# The DESI Single Fiber Lens Search. I. Four Thousand Spectroscopically Selected Galaxy–Galaxy Gravitational Lens Candidates

Juliana S. M. Karp[1,2,3], David J. Schlegel[3], Xiaosheng Huang[3,4], Nikhil Padmanabhan[5], Adam S. Bolton[6,7], Christopher J. Storfer[8], J. Aguilar[3], S. Ahlen[9], S. Bailey[3], D. Bianchi[10,11], D. Brooks[12], F. J. Castander[13,14], T. Claybaugh[3], A. Cuceu[3], A. de la Macorra[15], J. Della Costa[7,16], P. Doel[12], A. Font-Ribera[17], J. E. Forero-Romero[18,19], E. Gaztañaga[13,14,20], S. Gontcho A Gontcho[3,21], G. Gutierrez[22], K. Honscheid[23,24,25], M. Ishak[26], J. Jimenez[17], R. Joyce[7], S. Juneau[7], D. Kirkby[27], A. Kremin[3], C. Lamman[25], M. Landriau[3], L. Le Guillou[28], M. Manera[17,29], P. Martini[23,30], A. Meisner[7], R. Miquel[17,31], J. Moustakas[32], S. Nadathur[20], W. J. Percival[33,34,35], C. Poppett[3,36,37], F. Prada[38], I. Pérez-Ràfols[39], G. Rossi[40], E. Sanchez[41], M. Schubnell[42], D. Sprayberry[7], G. Tarlé[43], B. A. Weaver[7], R. Zhou[3,44] and The DESI Collaboration

[1] Department of Astronomy, University of Washington, 3910 15th Avenue NE, Seattle, WA 98195, USA; jsmkarp@uw.edu
[2] Department of Astronomy, Yale University, 219 Prospect Street, New Haven, CT 06511, USA
[3] Lawrence Berkeley National Laboratory, 1 Cyclotron Road, Berkeley, CA 94720, USA
[4] Department of Physics & Astronomy, University of San Francisco, San Francisco, CA 94117, USA
[5] Department of Physics, Yale University, New Haven, CT 06511, USA
[6] SLAC National Accelerator Laboratory, Menlo Park, CA 94025, USA
[7] NSF NOIRLab, 950 N. Cherry Avenue, Tucson, AZ 85719, USA
[8] Institute for Astronomy, University of Hawaii, Honolulu, HI 96822-1897, USA
[9] Department of Physics, Boston University, 590 Commonwealth Avenue, Boston, MA 02215, USA
[10] Dipartimento di Fisica "Aldo Pontremoli," Università degli Studi di Milano, Via Celoria 16, I-20133 Milano, Italy
[11] INAF—Osservatorio Astronomico di Brera, Via Brera 28, 20122 Milano, Italy
[12] Department of Physics & Astronomy, University College London, Gower Street, London, WC1E 6BT, UK
[13] Institut d'Estudis Espacials de Catalunya (IEEC), c/ Esteve Terradas 1, Edifici RDIT, Campus PMT-UPC, 08860 Castelldefels, Spain
[14] Institute of Space Sciences, ICE-CSIC, Campus UAB, Carrer de Can Magrans s/n, 08913, Bellaterra, Barcelona, Spain
[15] Instituto de Física, Universidad Nacional Autónoma de México, Circuito de la Investigación Científica, Ciudad Universitaria, Cd. de México C. P. 04510, México
[16] Department of Astronomy, San Diego State University, 5500 Campanile Drive, San Diego, CA 92182, USA
[17] Institut de Física d'Altes Energies (IFAE), The Barcelona Institute of Science and Technology, Edifici Cn, Campus UAB, 08193, Bellaterra (Barcelona), Spain
[18] Departamento de Física, Universidad de los Andes, Cra. 1 No. 18A-10, Edificio Ip, CP 111711, Bogotá, Colombia
[19] Observatorio Astronómico, Universidad de los Andes, Cra. 1 No. 18A-10, Edificio H, CP 111711 Bogotá, Colombia
[20] Institute of Cosmology and Gravitation, University of Portsmouth, Dennis Sciama Building, Portsmouth, PO1 3FX, UK
[21] University of Virginia, Department of Astronomy, Charlottesville, VA 22904, USA
[22] Fermi National Accelerator Laboratory, PO Box 500, Batavia, IL 60510, USA
[23] Center for Cosmology and AstroParticle Physics, The Ohio State University, 191 West Woodruff Avenue, Columbus, OH 43210, USA
[24] Department of Physics, The Ohio State University, 191 West Woodruff Avenue, Columbus, OH 43210, USA
[25] The Ohio State University, Columbus, 43210 OH, USA
[26] Department of Physics, The University of Texas at Dallas, 800 W. Campbell Road, Richardson, TX 75080, USA
[27] Department of Physics and Astronomy, University of California, IR 92697, USA
[28] Sorbonne Université, CNRS/IN2P3, Laboratoire de Physique Nucléaire et de Hautes Energies (LPNHE), FR-75005 Paris, France
[29] Departament de Física, Serra Húnter, Universitat Autònoma de Barcelona, 08193, Bellaterra (Barcelona), Spain
[30] Department of Astronomy, The Ohio State University, 4055 McPherson Laboratory, 140 W 18th Avenue, Columbus, OH 43210, USA
[31] Institució Catalana de Recerca i Estudis Avançats, Passeig de Lluís Companys, 23, 08010 Barcelona, Spain
[32] Department of Physics and Astronomy, Siena University, 515 Loudon Road, Loudonville, NY 12211, USA
[33] Department of Physics and Astronomy, University of Waterloo, 200 University Avenue W, Waterloo, ON N2L 3G1, Canada
[34] Perimeter Institute for Theoretical Physics, 31 Caroline Street North, Waterloo, ON N2L 2Y5, Canada
[35] Waterloo Centre for Astrophysics, University of Waterloo, 200 University Avenue W, Waterloo, ON N2L 3G1, Canada
[36] Space Sciences Laboratory, University of California, Berkeley, 7 Gauss Way, Berkeley, CA 94720, USA
[37] University of California, Berkeley, 110 Sproul Hall #5800, Berkeley, CA 94720, USA
[38] Instituto de Astrofísica de Andalucía (CSIC), Glorieta de la Astronomía, s/n, E-18008 Granada, Spain
[39] Departament de Física, EEBE, Universitat Politècnica de Catalunya, c/Eduard Maristany 10, 08930 Barcelona, Spain
[40] Department of Physics and Astronomy, Sejong University, 209 Neungdong-ro, Gwangjin-gu, Seoul 05006, Republic of Korea
[41] CIEMAT, Avenida Complutense 40, E-28040 Madrid, Spain
[42] Department of Physics, University of Michigan, 450 Church Street, Ann Arbor, MI 48109, USA
[43] University of Michigan, 500 S. State Street, Ann Arbor, MI 48109, USA
[44] National Astronomical Observatories, Chinese Academy of Sciences, A20 Datun Road, Chaoyang District, Beijing 100101, People's Republic of China

## Abstract

We present 4110 strong gravitational lens candidates, 3887 of which are new discoveries, selected from a sample of 5,837,154 luminous red galaxies (LRGs) observed with the Dark Energy Spectroscopic Instrument (DESI). Candidates are identified via the presence of background ionized oxygen [O II] nebular emission lines in the foreground LRG spectra, which may originate from the lensing of higher-redshift star-forming galaxies. Using the measured foreground redshift, background redshift, and integrated flux of the background [O II] doublet, we integrate over impact parameters to compute the probability that each candidate is a lens. We expect 53% of candidates to be true lenses with Einstein radii ranging from 0.″1–4″, which can be confirmed with high-resolution imaging. Confirmed strong lenses from this sample will form a valuable cosmological data set, as strong gravitational lensing is the only method to directly measure dark matter halo substructure at cosmological distances. We independently recover the host of the multiply imaged gravitationally lensed type Ia supernova iPTF16geu. Monitoring these lenses for future multiply lensed transients will enable (a) $H_0$ measurements via time-delay cosmography and (b) substructure measurements via flux ratios.

## 1. Introduction

Strong gravitational lensing (A. Einstein 1936) occurs when the trajectory of light is bent by a large foreground mass, creating multiple distorted images of the background source and magnifying its light. Strong lensing is a powerful tool to study cosmology. It can reveal galaxies' and galaxy clusters' central dark matter (DM) profiles (e.g., C. S. Kochanek 1991; D. J. Sand et al. 2002; T. Treu & L. Koopmans 2002; A. S. Bolton et al. 2006, 2008; M. Bradač et al. 2008; X. Huang et al. 2009; E. Jullo et al. 2010; C. Grillo et al. 2015; Y. Shu et al. 2017), helping to resolve the cusp-core controversy, a point of contention between cold dark matter (CDM) and other DM models. Lensing can also reveal DM substructures within galaxy and cluster halos at cosmological distances (e.g., C. S. Kochanek & N. Dalal 2004; L. V. E. Koopmans 2005; M. J. Jee et al. 2007; S. Vegetti et al. 2010, 2012). Galaxy–galaxy lenses are particularly useful for constraining the mass distribution of the foreground lens, since, unlike in the case of lensed quasars, the background lensed source is spatially resolved. Because strong lensing magnifies light from the background source, it also allows for unique opportunities to study faint objects at high redshifts, which cannot otherwise be observed. Additionally, time-delay measurements from multiply imaged lensed supernovae (e.g., J. D. R. Pierel & S. Rodney 2019; S. A. Rodney et al. 2021; P. L. Kelly et al. 2023; B. L. Frye et al. 2024; X. Li & K. Liao 2024; J. D. R. Pierel et al. 2024; S. H. Suyu et al. 2024), quasars (e.g., S. Birrer et al. 2020; A. J. Shajib et al. 2020; K. C. Wong et al. 2020; L.-F. Wang et al. 2022), and gamma-ray bursts (e.g., X. Ding et al. 2021; S.-S. Du et al. 2023) provide an independent method to constrain $H_0$ competitive with other cosmological probes (S. Refsdal 1964; E. V. Linder 2011; T. Treu & P. J. Marshall 2016). Future large-sky surveys like the Vera C. Rubin Observatory's Legacy Survey of Space and Time (LSST; Ž. Ivezić et al. 2019) and those conducted by Euclid (R. Laureijs et al. 2011; Euclid Collaboration et al. 2025a) and the Nancy Grace Roman Space Telescope (D. Spergel et al. 2015) will monitor catalogs of known strong lenses for such transients (J. R. Pierel & S. Rodney 2020). Increasing the number of known strong gravitational lensing systems is therefore paramount to addressing many open questions in cosmology.

Currently, there exist $\mathcal{O}(10^4)$ strong lensing candidates and $\sim$400 confirmed strong lensing systems. To confirm a strong lensing system, both high-resolution imaging and spectroscopic observations are required. It is necessary to resolve the multiply imaged background sources and to confirm that the source and lens are at different redshifts. Some candidates are initially identified in ground-based imaging surveys (e.g., A. More et al. 2012; C. Jacobs et al. 2017; X. Huang et al. 2020, 2021; C. Storfer et al. 2022), then followed up with high-resolution imaging and spectroscopy. Others are initially discovered spectroscopically (e.g., A. S. Bolton et al. 2004, 2006, 2008; J. R. Brownstein et al. 2012; M. S. Talbot et al. 2018, 2021), then followed up with high-resolution imaging for confirmation. The spectroscopic identification method, which we employ in this work, was first proposed by S. J. Warren et al. (1996) and further developed by A. S. Bolton et al. (2004). It consists of searching for spectral signatures (e.g., nebular emission lines) from a higher-redshift galaxy in superposition with the foreground target's spectrum. A massive galaxy should effectively lens light from any sufficiently distant background object at small enough impact parameter, so these background emission lines are indications of lensing. Spectroscopic lens searches allow us to determine redshifts for both the foreground lens and background source, which is not possible when identifying systems in imaging surveys.

Imaging and spectroscopic lens discovery methods are complementary, in that they are most successful at identifying different populations of strong lenses. Only systems with Einstein radii large enough for an instrument to resolve both the lens and the source can be identified in ground-based imaging. On the other hand, strong lenses that are initially identified in spectroscopic surveys must have smaller Einstein radii, because light from both the lens and the source must be detected within one fiber. The Einstein radius is dependent on the masses and redshifts of the source and lens (T. Treu 2010).

Recently, a third lens selection method has been developed. This method identifies gravitational lenses via a pair-wise search of spectroscopic fibers with small angular separations between them (Y.-M. Hsu et al. 2025). The pair-wise search preferentially selects lenses with higher Einstein radii than those found through single-fiber spectroscopic searches, but smaller than imaging-based searches.

## 2. Data and Search Sample

The Dark Energy Spectroscopic Instrument (DESI; DESI Collaboration et al. 2016a, 2016b) is a 5000-fiber spectrograph with a 3.°2 diameter field of view (DESI Collaboration et al. 2022) mounted atop the 4-meter Mayall Telescope at the Kitt Peak National Observatory in Arizona. DESI is creating the densest map yet of the local Universe by observing over 50 million galaxies and quasars out to redshift $\sim$2, with the goal of measuring baryon acoustic oscillations and large-scale structure growth via redshift-space distortions.

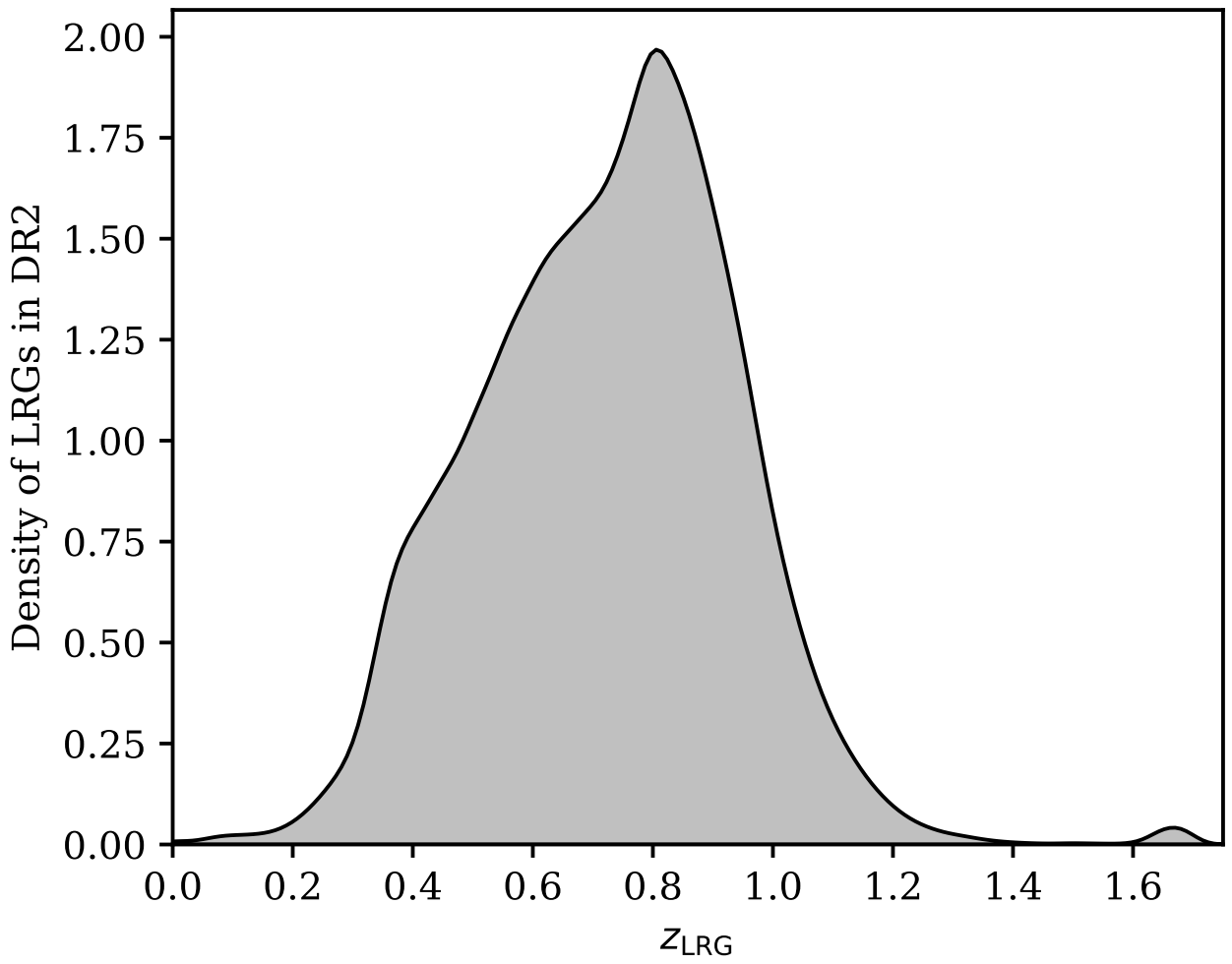

**Figure 1.** Distribution of luminous red galaxy (LRG) redshifts ($z_{\mathrm{LRG}}$) observed in dark time included in the DESI Data Release 2 (DR2). Densities are calculated using a Gaussian density kernel estimator with a bandwidth factor of 0.1. We search this sample of 5,837,154 spectra for lenses.

DESI observes continuously between $\lambda = 3600$ and 9800 Å with resolution $\lambda/\Delta\lambda = 2200$–5000, increasing with wavelength. DESI is a successor to the Sloan Digital Sky Survey's (SDSS; D. G. York et al. 2000) Baryon Oscillation Spectroscopic Survey (BOSS; K. S. Dawson et al. 2013). Before DESI, the largest three-dimensional map of the Universe was constructed by SDSS, which observed $\sim$4 million extragalactic sources with spectral resolution $\lambda/\Delta\lambda = 1800$. The spectroscopic lens search technique was pioneered by A. S. Bolton et al. (2004) for SDSS data. DESI is observing an order of magnitude more sources with better spectral resolution, making it an ideal survey to search for new spectroscopic lenses.

We limit our strong lens search to luminous red galaxies (LRGs). LRGs are massive, bright, and relatively quiescent early-type galaxies. Due to their high masses and intermediate redshifts ($z \sim 0.3$–1.2; see Figure 1), LRGs have high chances of being lenses to objects at higher redshifts. Their spectral energy distributions are well defined with a characteristic break at 4000 Å (N. Padmanabhan et al. 2005, 2007), meaning background lensed sources can be identified with high confidence via deviations from the known standard LRG spectrum. Additionally, as LRGs are expected to have little dust, extinction of emission from any lensed background sources should be minimal.

DESI spectroscopic targeting is based on the DESI Legacy Imaging Surveys (DESI LS; A. Dey et al. 2019), which is itself composed of several photometric surveys: the Dark Energy Camera Legacy Survey (DECaLS) conducted by the Dark Energy Camera (B. Flaugher et al. 2015) on the Blanco 4-meter telescope, the Beijing-Arizona Sky Survey (BASS) conducted by the 90Prime Camera (G. G. Williams et al. 2004) on the Bock 2.3 meter telescope, and the Mayall $z$-band Legacy Survey (MzLS) conducted by the Mosaic3 camera (A. Dey et al. 2016) on the Mayall 4 meter telescope. The DESI LS Data Release 10 (DR10) also includes four infrared bands (3.4, 4.6, 12, and 22 $\mu$m) from the unWISE survey (E. F. Schlafly et al. 2019) conducted by the Near-Earth Object Wide-field Infrared Survey Explorer (NEOWISE; A. Mainzer et al. 2011). DESI's LRG targeting strategy is described in detail in R. Zhou et al. (2023). In brief, DESI targets LRGs at a rate of 350 per square degree up to $z = 1.0$, such that they are a dominant portion of the observed sample at intermediate redshifts. In comparison, SDSS targeted LRGs at a rate of 119 per square degree at $z \leqslant 0.6$. At $z < 0.6$, DESI selects LRGs using the same cuts as BOSS, described in detail in D. J. Eisenstein et al. (2001). At $z > 0.6$, the 4000 Å break is redshifted beyond the end of the $r$ band. At these redshifts, DESI uses WISE photometry (E. L. Wright et al. 2010) to select LRGs via cuts in W1 $-$ $r$ and $r - z$ color (DESI Collaboration et al. 2016a).

We conduct our search on the combined Main Survey (main) and Survey Validation 3 (SV3) samples of the DESI Year 3 internal data release, also named the spectroscopic production *Loa*, and forthcoming as Data Release 2 (DR2). *Loa* main includes a total of 53,937,655 spectra collected across both dark and bright time between 2021 May 15 and 2024 April 10. SV3 includes 1,598,831 spectra observed across bright and dark time between 2021 April 6 and 2021 June 11. Among these, we choose only spectroscopically confirmed galaxies ("SPECTYPE" = "GALAXY") targeted as LRGs ("DESITARGET" and desimask["LRG"] $\neq$ 0) observed in dark time ("PROGRAM" = "dark"). We choose to only consider LRGs observed in dark time to maximize the signal-to-noise ratio of the faint background emission features we target. These cuts yield a search sample of 5,698,659 spectra in DESI main and 138,495 in SV3, for a grand total of 5,837,154 LRG spectra. Extrapolating from previous spectroscopic identification lens surveys, we expect that roughly 1 in 1000 LRGs will be strong lenses (A. S. Bolton et al. 2006) to galaxies at higher redshifts. From this assumption, we expect to identify $\sim 10^3$ strong gravitational lenses in this study.

## 3. Candidate Selection Methods

We employ largely the same spectroscopic lens selection method as A. S. Bolton et al. (2004), with some modifications. For each object, we subtract the best-fit galaxy model spectrum from the sky-subtracted data, then search the residual spectrum for [O II] $\lambda\lambda$3727 emission indicative of a star-forming background galaxy.

The REDROCK[45] (S. Bailey et al. 2026, in preparation) component of the DESI spectroscopic pipeline (J. Guy et al. 2023) fits each spectrum $f_\lambda$ to a linear combination of principal component analysis (PCA) template spectra for quasars, galaxies, and stars. REDROCK returns both the best-fit object type and redshift. The best-fit models $\tilde{f}_\lambda$ are not directly stored by the pipeline, but they can be reconstructed from the recorded spectrum type, best-fit redshift, and PCA coefficients, which are stored in the "SPECTYPE," "Z," and "COEFF" columns of the redshift catalog, respectively. To do this, we load the eigentemplates, redshift the template wavelength to the REDROCK-determined best-fit redshift, and take the dot product of the best-fit coefficients and the templates corresponding to the best-fit spectrum type. We then interpolate the reconstructed best-fit model spectrum $\tilde{f}_\lambda$ to match the wavelength sampling of the data $\lambda$ and take their difference to obtain the residual spectrum $f_{\lambda,r}$.

$$f_{\lambda,r} = f_\lambda - \tilde{f}_\lambda. \quad (1)$$

Unusually strong or broad emission lines in the foreground galaxy's spectrum that are not perfectly fit by REDROCK can

[45] https://github.com/desihub/redrock

**Table 1**
Description of Common Emission Lines Which May Not Be Perfectly Fit by REDROCK in the Foreground Galaxy's Spectrum

| Line | Wavelength (Å) | Mask (km $s^{-1}$) |
|---|---|---|
| Ly$\alpha$ | 1215.67 | 300 |
| N V | 1240.81 | 300 |
| C IV | 1549.48 | 300 |
| He II | 1640.42 | 300 |
| C III] | 1908.73 | 300 |
| Mg II | 2800.32 | 300 |
| [O II] | 3729.09 | 300 |
| [O II] | 3729.88 | 2000 |
| [Ne III] | 3869.86 | 2000 |
| H$\gamma$ | 4341.68 | 300 |
| [O III] | 4364.44 | 300 |
| H$\beta$ | 4862.68 | 300 |
| [O III] | 4960.29 | 3000 |
| [O III] | 5008.24 | 4000 |
| [O I] | 5578.89 | 300 |
| [N II] | 5756.19 | 3000 |
| [O I] | 6302.05 | 300 |
| [O I] | 6365.54 | 300 |
| [N II] | 6549.86 | 300 |
| H$\alpha$ | 6564.61 | 300 |
| [N II] | 6585.27 | 300 |
| [S II] | 6718.29 | 300 |
| [S II] | 6732.68 | 300 |

**Note.** All wavelengths are given in rest-frame vacuum wavelengths. We apply masks around these foreground lines by setting the inverse variance to 0 within the reported full width for each line (column 3) centered on the line wavelength (column 2). This reduces the number of false-positive lens candidates selected by the pipeline.

contaminate the calculated residual spectrum and increase the number of false-positive lens candidates. We address this by masking out values around common emission-line wavelengths at the REDROCK best-fit foreground redshift. These specific lines, along with the width of the mask applied in each case, are described in Table 1. We mask these portions of the spectrum by setting the inverse variance $1/\sigma^2$ (with $\sigma$ being the uncorrelated Gaussian error) to 0 within the reported mask width around the line's central wavelength. Most mask widths are small, but several lines that are commonly known to have broad wings that are not well fit by REDROCK require wider masks (e.g., certain [Ne III], [O III], and [N II] lines). We also mask out the area within $\pm$ 150 km $s^{-1}$ of the 5577 Å sky line, which is necessary due to the sky line's brightness and the inconsistent quality of its REDROCK fit. In total, the applied masks cover a width of 19,000 km $s^{-1}$, corresponding to, at most, ~6% of the wavelength range we search for background [O II] emission lines. Given the vast number of LRGs in the search sample, this explicit masking only results in a small loss in our final number of candidates. However, in the future, it may be advantageous to instead implement the noise-rescaling process employed by A. S. Bolton et al. (2004, as described in their Appendix A) to preserve full spectral search coverage while still minimizing false positives from poorly fit foreground and sky emission lines.

The best-fit REDROCK model accounts for continuum emission. If there is no secondary object within the LRG fiber, the masked residual spectrum should be consistent with zero at every wavelength (within uncorrelated Gaussian noise). Therefore, any feature in the residual spectrum is indicative of another source in superposition along the line of sight. Given the small size of a DESI fiber (0."75 in radius), the impact parameter of this source compared to the foreground LRG must be small. If the redshift of the background galaxy ($z_{BG}$) is significantly higher than the redshift of the foreground LRG ($z_{FG}$), it is likely the LRG is acting as a lens to this background source. Generally, we expect lenses to obey $z_{BG} \sim 2z_{FG}$, but this can vary widely.

The [O II] $\lambda\lambda$3727 doublet in particular is a strong tracer of star formation. We expect many galaxies in the background redshift range that has the highest likelihood of being lensed by DESI LRGs to be star-forming. The [O II] $\lambda\lambda$3727 line is distinct, as it is a doublet, and therefore can be used to uniquely determine the redshift of the lensed source without requiring a more computationally expensive multiple line search. DESI is able to observe [O II] $\lambda\lambda$3727 up to $z \sim 1.6$. For all of these reasons, we choose to target the [O II] $\lambda\lambda$3727 doublet in our residual spectrum emission-line search. The search pipeline consists of three main phases: (1) an initial linear least-squares fit to the residual spectra to identify systems of interest, (2) a secondary nonlinear Markov Chain Monte Carlo (MCMC) fit to these systems of interest at the wavelengths of the previously identified background features of interest, and (3) a final set of cuts to remove remaining contaminants. We explain these phases in greater detail.

*Phase 1.* We create a mock double-Gaussian profile $G(\lambda, z)$ that is zero everywhere except around the central wavelengths of the [O II] $\lambda\lambda$3727 doublet in rest-frame vacuum units ($\lambda_{1,z=0} = 3727.092$ Å and $\lambda_{2,z=0} = 3729.875$ Å) (National Institute of Standards and Technology 2024).[46]

$$G(\lambda, z) = \sum_{i=1}^{2} a_i \frac{1}{\sigma_{i,z}\sqrt{2\pi}} \exp\left(-\frac{1}{2}\frac{(\lambda - \lambda_{i,z})^2}{\sigma_{i,z}^2}\right). \quad (2)$$

We require a fixed line flux ratio of 1 to 1.3 (such that $a_1 = a$ and $a_2 = 1.3a$), allowing the doublet's amplitude $a$ to be fit as a free parameter. The 1:1.3 [O II] $\lambda\lambda$3727 flux ratio corresponds to an electron density of $n_e = 100\,\mathrm{cm}^{-1}$, found by the MOSFIRE Deep Evolution Field (MOSDEF) survey (M. Kriek et al. 2015) to be a typical value for star-forming galaxies around $z \sim 2$ (R. L. Sanders et al. 2016). We fit this model to the residual spectrum by searching over a two-dimensional grid of emission-line widths $\sigma_i$ and redshifts $z$. The searched widths correspond to line-of-sight velocity dispersions $v_{disp} = 50,\ 75,$ and 100 km $s^{-1}$, and the redshift grid spans from $z = 0.3$ to $z = 1.50$ in increments of $10^{-4}$ in redshift space, corresponding to 30 km $s^{-1}$. Although DESI can detect [O II] emission up to $z \sim 1.6$, we choose to limit the search range to background redshifts of $z_{BG} < 1.50$, where the DESI spectroscopic throughput is high, the atmospheric emission lines are not as prominent, and the search is expected to be more successful. At each position in this grid, we recalculate $G(\lambda, z)$ by redshifting the central wavelengths of the [O II] $\lambda\lambda$3727 doublet

$$\lambda_{i,z} = \lambda_{i,z=0} \cdot (1 + z) \quad (3)$$

[46] https://www.nist.gov/pml/atomic-spectra-database

and translating the searched velocity dispersions to wavelength-space widths according to this redshift.

$$\sigma_{i,z} = \lambda_{i,z} \cdot \frac{v_{\rm disp}}{c}. \quad (4)$$

At every redshift-velocity dispersion point in the search grid, we perform a least-squares fit between the residual spectrum $f_{\lambda,r}$ and the mock double-Gaussian profile $G(\lambda, z)$. We store the best-fit line height coefficient, the error on this coefficient, and the reduced chi squared ($\chi_r^2$) of the fit. For each spectrum, we then record the redshift corresponding to the minimum $\chi_r^2$ as the best-fit background redshift. We estimate our confidence in the best-fit model by calculating $\Delta\chi_r^2$, or the difference in $\chi_r^2$ between the best redshift model and the second-best model at a significantly different redshift. We also record the difference in $\chi_r^2$ between the best redshift model and no model to estimate the significance of the [O II] detection. As a first cut, we select all spectra for which (a) the $\chi_r^2$ value is lowered by at least 100 by the presence of the [O II] emission-line model when compared to no model, (b) the best-fit background [O II] doublet amplitude $a$ is positive, and (c) the background [O II] doublet's redshift is greater than the foreground LRG's REDROCK-fit redshift. This yields 55,258 spectra of interest, 54,476 of which are in main and 782 of which are in SV3.

*Phase 2.* We then refit these 55,258 spectra of interest with a secondary nonlinear model to refine the best-fit parameters and increase the purity of our sample. We create another mock double-Gaussian profile $G(\lambda)$, this time allowing the velocity dispersion $v_{\rm disp}$ and two line amplitudes $a_1$ and $a_2$ to all be fit independently as free parameters. We fix the redshift of the background emission feature at the best-fit redshift identified by the least-squares fit $z_{\rm BG}$, such that redshift is no longer a free parameter. To best sample parameter space and avoid falling into local minima as a gradient descent algorithm might, we use the MCMC algorithm implemented in the EMCEE Python package (D. Foreman-Mackey et al. 2013). We set uniform priors between $-50 < a_i < 50$ and $0 < v_{\rm disp} < 2000$ for our three free parameters. We choose not to force $a_i > 0$ to later be able to remove unphysical detections. For each spectrum, we initialize 32 walkers with starting positions $a_i = 1$ and $v_{\rm disp} = 50$. We run each EMCEE walker for a total of 1000 steps, which we determine through trial and error is sufficient for the MCMC to converge. From the posterior distribution of each residual spectrum, we recover the converged values for $a_1$, $a_2$, and $v_{\rm disp}$ at $z = z_{\rm BG}$, as well as the $1\sigma$ errors on these parameters. We also calculate both the $\chi_r^2$ of the fit within $\pm$ 15 Å of the center of the background emission feature of interest, which we call the MCMC doublet $\chi_r^2$, and the per-pixel signal-to-noise ratio of the doublet. We use these parameters to make a second set of cuts on our spectra of interest.

From the 55,258 spectra modeled with MCMC, we keep only those that satisfy:

1. the amplitudes of both recovered lines are positive ($a_1 > 0$ and $a_2 > 0$);
2. the ratio between the two best-fit line amplitudes is <2 ($0.5 < a_1/a_2 < 2$). These specific bounds are based on both physically motivated analyses (L. Simard 1996)[47] and empirical studies (A. Harshan et al. 2020) of [O II] $\lambda\lambda$3727 emission in galaxies at $z \sim 2$;
3. the MCMC doublet $\chi_r^2$ is low ($0 < \chi_r^2 < 3$);
4. the best-fit velocity dispersion is physically plausible for star-forming galaxies in our recovered background redshift range ($20\,{\rm km\,s^{-1}} < v_{\rm disp} < 400\,{\rm km\,s^{-1}}$);
5. the doublet's per-pixel signal-to-noise ratio is >1;
6. none of the data points within $\pm$ 15 Å of the center of the doublet are masked out as potential foreground emission line or sky line contaminants.

This yields a sample of 4505 remaining spectra of interest (4378 in main and 127 in SV3).

*Phase 3.* We remove several obvious remaining contamination sources. These contamination sources appear visually as overdense linear streaks in background [O II] doublet redshift $z_{\rm BG}$ versus REDROCK-identified foreground LRG redshift $z_{\rm FG}$ space (central panel of Figure 2). In such redshift–redshift space, an unfit, unmasked emission line in the foreground spectrum that passes our prior cuts would lie along the trace of a linear function with an intercept of zero in $1 + z_{\rm BG}$ versus $1 + z_{\rm FG}$ space. Translating this to $z_{\rm BG}$ versus $z_{\rm FG}$ space, we find that all of the streak-like features, aside from those clustered around $z_{\rm FG} = 0$ (highlighted in pink), obey this behavior. Calculating the rest-frame wavelength of the foreground emission line responsible for each contaminating streak is then simple:

$$\lambda_{\rm em} = \frac{1 + z_{\rm BG}}{1 + z_{\rm FG}} \lambda_{\rm [OII]} \quad (5)$$

where $\lambda_{\rm [O\,II]}$ is the central rest-frame wavelength of the [O II] $\lambda\lambda$3727 doublet. By comparing the recovered $\lambda_{\rm em}$ wavelengths to a table of common galaxy emission lines (Chojnowski 2025),[48] we find that five of these streak-like features can be explained by the following rare emission lines in the foreground LRG's spectrum being badly fit or entirely unfit by REDROCK and appearing in the residual spectrum when they should not: He II $\lambda$4686, [Fe VII] $\lambda$5159, [Fe VII] $\lambda$6087, Ca II $\lambda$8542, and Ca II $\lambda$8662. We remove all spectra that lie within $\pm$0.003 of these lines in redshift–redshift space, as highlighted in Figure 2. The feature highlighted in green is consistent with an emission line with a rest-frame wavelength of 6326 Å, but this does not correspond to any common galactic emission line listed in the emission-line reference table. Despite its origin being unknown, we remove the spectra associated with this feature (within $\pm$0.003 in redshift–redshift space) from the catalog too, as they are clearly the result of a contaminant of some sort. Finally, we remove sources with $z_{\rm FG} < 0.005$—those highlighted in pink—as they are likely not LRGs but rather M stars within our galaxy whose light is blended along the line of sight with that from a faint background galaxy. After this final round of cuts, we are left with 4110 sources, which comprise the final catalog of strong gravitational lens candidates.

## 4. The Lens Candidate Catalog

The catalog is composed of 4110 strong galaxy–galaxy gravitational lens candidates selected from the sample of 5,698,659 LRGs in the DESI Year 3 internal data release *Loa*, forthcoming as DESI DR2. The candidates all present background [O II] $\lambda\lambda3727$ emission at a redshift higher than that of the foreground target LRG. They have been identified via a three-phase spectroscopic selection process, described in detail in Section 3, designed to minimize false positives from unfit foreground emission lines. The candidates' distribution in redshift–redshift space is shown in Figure 2. Foreground LRG redshifts span from $z_{FG} = 0.01$ to $z_{FG} = 1.40$, while background [O II] redshifts range from $z_{BG} = 0.33$ to $z = 1.50$. The full catalog is publicly available at https://data.desi.lbl.gov/public/papers/gqp/single-fiber-lens/. We present an abridged version in Table 2.

[47] https://www.ucolick.org/~simard/phd/root/node21.html

[48] astronomy.nmsu.edu/drewski/tableofemissionlines.html

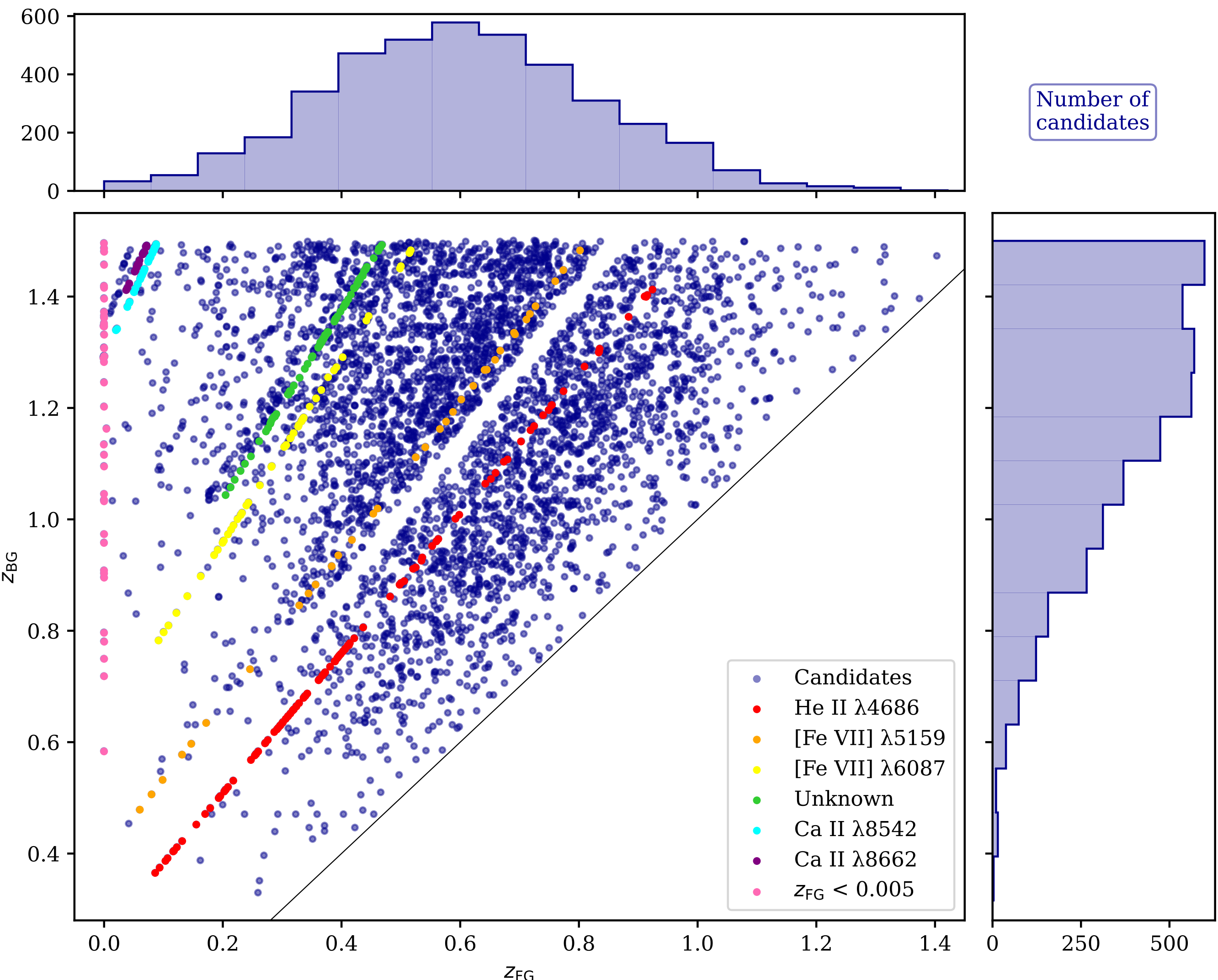

**Figure 2.** Distribution of foreground (REDROCK best fit) and background (determined by the secondary [O II] doublet) redshifts for lens candidates. The solid black line indicates where $z_{FG} = z_{BG}$. Spectra of interest remaining after the secondary MCMC fit and cuts are applied (end of Phase 2). Several linear streak-like features are apparent in the $z_{BG}$ vs. $z_{FG}$ distribution. These contaminants are largely due to rare emission lines in the foreground LRG spectrum that are poorly fit by REDROCK, as indicated by their colors and labels. Sources with $z_{FG} < 0.005$ are likely M stars blended with faint galaxies. We remove these contaminants (Phase 3). The remaining 4110 sources plotted here form the lens candidate catalog (dark blue). The side panels show histogram distributions of the 4110 candidates' foreground and background redshifts.

In Figure 3, we plot spectra and imaging cutouts of 20 example lens candidates, selected at random from the full catalog of 4110 candidates.[49] We show the full spectrum, a zoom on the recovered background [O II] $\lambda\lambda3727$ doublet, and imaging of the source from the DESI LS DR10. For many of our candidates, DESI LS lacks the angular resolution necessary to resolve the source of the detected background [O II] doublet for lens confirmation. We discuss various alternate possibilities for high-resolution imaging confirmation in Section 6.2.

To better characterize the background source sample, we investigate the distribution of best-fit parameters for the recovered background [O II] $\lambda\lambda3727$ doublets (Figure 4). In the top panel, we bin the candidates by the redshift of the background [O II] doublet $z_{BG}$ identified by the initial linear least-squares fitting algorithm. The recovered $z_{BG}$ span from 0.33–1.50, at which point the [O II] doublet is redshifted beyond the wavelength range that DESI can observe with high confidence. The number of background sources increases monotonically from $z_{BG} = 0.33$ to $z_{BG} \sim 1.15$, then plateaus in the range $\sim 1.15 \leqslant z_{BG} \leqslant 1.50$. The monotonic increase is expected because the lensing kernel increases with increasing redshift separation between the foreground lensing galaxy and

[49] Similar plots of all 4110 candidates are available in Zenodo at DOI:10.5281/zenodo.17795798.

**Table 2**
A Subsample of Candidates and Their Characteristics

| Name | R.A. (deg) | decl. (deg) | $z_{FG}$ | $z_{BG}$ | $\theta_E$ (arcsec) | $p_L$ | $a_1$ ($10^{-17}$ erg s$^{-1}$ cm$^{-2}$) | $a_2$ ($10^{-17}$ erg s$^{-1}$ cm$^{-2}$) |
|---|---|---|---|---|---|---|---|---|
| DESI-128.9593+40.7310 | 128.9593 | 40.7310 | 0.6299 | 1.2084 | 0.8754 | 0.6857 | 4.4501 | 4.5637 |
| DESI-129.3791+41.5529 | 129.3791 | 41.5529 | 0.5616 | 0.8377 | 0.4989 | 0.1209 | 3.3390 | 3.9388 |
| DESI-118.0223+36.3576 | 118.0223 | 36.3576 | 0.3376 | 0.7661 | 0.7283 | 0.5202 | 5.9247 | 8.1157 |
| DESI-118.5190+36.3685 | 118.5190 | 36.3685 | 1.3137 | 1.4245 | 0.0930 | 0.0047 | 5.6784 | 9.5490 |
| DESI-275.9965+50.6586 | 275.9965 | 50.6586 | 0.5642 | 1.3364 | 1.3089 | 0.9987 | 3.1349 | 5.2845 |
| DESI-232.8583+15.1483 | 232.8583 | 15.1483 | 0.9947 | 1.3901 | 0.3728 | 0.0459 | 2.9461 | 3.1684 |
| DESI-039.6386-09.2392 | 39.6386 | −9.2392 | 0.5308 | 1.3476 | 1.4949 | 1.0000 | 5.0879 | 6.8906 |
| DESI-134.6454-05.1633 | 134.6454 | −5.1633 | 0.9879 | 1.3088 | 0.3225 | 0.0252 | 2.3736 | 3.6986 |
| DESI-266.1728+59.4211 | 266.1728 | 59.4211 | 0.5298 | 1.4678 | 0.8804 | 0.9230 | 6.2528 | 8.1012 |
| DESI-263.8834+35.4678 | 263.8834 | 35.4678 | 0.5509 | 1.4488 | 1.7131 | 0.9913 | 3.9680 | 3.9066 |
| DESI-026.5159-16.2704 | 26.5159 | −16.2704 | 0.8287 | 1.3360 | 1.5263 | 0.9971 | 2.7757 | 3.6267 |
| DESI-228.0353-02.5792 | 228.0353 | −2.5792 | 0.7201 | 1.1980 | 1.1066 | 0.9397 | 2.8585 | 4.1149 |
| DESI-125.9734+34.1978 | 125.9734 | 34.1978 | 0.5094 | 0.8670 | 0.6343 | 0.2198 | 3.0252 | 4.5289 |
| DESI-168.6180+18.9102 | 168.6180 | 18.9102 | 0.5670 | 0.9378 | 1.2542 | 0.9989 | 6.1851 | 6.9058 |
| DESI-126.2741+31.6277 | 126.2741 | 31.6277 | 0.7660 | 1.2287 | 0.5306 | 0.2103 | 4.8588 | 5.7671 |
| DESI-125.6908+31.4574 | 125.6908 | 31.4574 | 0.6786 | 1.2905 | 0.6860 | 0.3631 | 4.1031 | 4.9669 |
| DESI-354.3825+26.5710 | 354.3825 | 26.5710 | 0.6960 | 1.4643 | 0.3910 | 0.0729 | 3.7677 | 4.2213 |
| DESI-051.3004-17.1041 | 51.3004 | −17.1041 | 0.4651 | 0.5857 | 0.4957 | 0.2070 | 5.3538 | 6.0410 |
| DESI-002.5400+05.5970 | 2.5400 | 5.5970 | 0.2380 | 1.0649 | 0.2628 | 0.0127 | 2.5668 | 3.4015 |
| DESI-045.4023-12.8390 | 45.4023 | −12.8390 | 0.9147 | 1.3640 | 0.4425 | 0.0838 | 3.0779 | 4.2690 |

**Note.** For each candidate, we report the system name, on-sky position, foreground lens redshift ($z_{FG}$), background source redshift ($z_{BG}$), Einstein radius ($\theta_E$, which we calculate in Section 6.1), lensing probability ($p_L$, a value between 0 and 1, also calculated in Section 6.1), and best-fit amplitudes for the 3727 Å and 3729 Å lines of the [O II] doublet ($a_1$ and $a_2$, respectively).

(This table is available in its entirety in machine-readable form in the online article.)

background source galaxy. Thus, for a sample of lenses with a uniform distribution of foreground redshifts, we expect to see a monotonic increase over $z_{BG}$ in the number of lens candidates found. The observed plateau in the distribution of background redshifts beyond $z_{BG} = 1.15$ can be explained by the balancing of three effects: (1) DESI only targets LRGs up to $z \sim 1$, which reduces the number of lenses with much greater background redshifts, (2) the incident flux from lensed sources at higher background redshifts will be lower and thus more difficult to detect, and (3) the lensing kernel increases at higher $z_{BG}$ due to the increase in difference between $z_{FG}$ and $z_{BG}$, which increases the number of lenses at higher $z_{BG}$.

In the middle panel, we bin the candidates by the doublet line flux ratio $a_1/a_2$. Note that $a_1$ and $a_2$ are the respective best-fit amplitudes of the 3726 Å and 3728 Å lines recovered by the secondary MCMC algorithm. The distribution spans the range $0.5 < a_1/a_2 < 2$, which represents hard cuts we imposed to limit candidates to the range of parameters that are physical for the [O II] doublet. As previously explained, these bounds are derived from both theoretical calculations and empirical studies of the environments of $z \sim 2$ galaxies. The distribution is not uniform within these bounds; rather, it peaks very strongly at $a_1/a_2 \sim 0.75$, corresponding to a line flux ratio of ∼1:1.3. This confirms the MOSDEF survey's (M. Kriek et al. 2015) finding that an electron density of $n_e = 100\,\mathrm{cm}^{-1}$ is a typical value for star-forming galaxies post–cosmic noon (R. L. Sanders et al. 2016).

In the bottom panel, we bin candidates by the velocity dispersion of the background [O II] doublet recovered by the secondary MCMC fit. This velocity dispersion is simply the best-fit width of the line translated from angstroms to kilometers per second according to the redshift, meaning it is the product of the instrumental and astrophysical dispersion. The DESI instrumental dispersion ranges from 45 km s$^{-1}$ in the blue to 20 km s$^{-1}$ in the red. Similarly to the line flux ratio distribution, the velocity dispersion distribution's bounds are hard cuts imposed during the selection process ($20\,\mathrm{km\,s^{-1}} < v_{disp} < 400\,\mathrm{km\,s^{-1}}$). The lower bound of 20 km s$^{-1}$ is chosen to eliminate cases where the background feature is not an emission line, but rather an artifact of the REDROCK fitting pipeline or a bad data point, which has been improperly masked. The upper bound of $v_{disp} < 400\,\mathrm{km\,s^{-1}}$ is chosen to eliminate extremely broad background emission features, which are likely not [O II] $\lambda\lambda$3727 doublets.

We find that the distribution of recovered background source velocity dispersions peaks at $v_{disp} \sim 80\,\mathrm{km\,s^{-1}}$, with over two-thirds (68.3%) of all systems having $v_{disp} < 100\,\mathrm{km\,s^{-1}}$. Only 7.7% of systems present background [O II] widths corresponding to velocity dispersions >200 km s$^{-1}$. This is a physically plausible background source velocity dispersion distribution. We expect most of our background emission sources to be spiral galaxies, which usually have velocity dispersions on the order of a couple hundred kilometers per second or less. For instance, the Milky Way, a relatively massive spiral galaxy at $z = 0$, has a rotational velocity of ∼220 km s$^{-1}$. The background galaxies we identify should be less massive than the Milky Way, given that they are at higher redshifts, and therefore have smaller velocity dispersions. Since DESI fibers are quite small, we are also likely only detecting light from portions of the background galaxy, which would contribute to a lower observed $v_{disp}$. Additionally, any rotating galaxy that is observed at an inclination angle that is not entirely edge-on will have a lower observed $v_{disp}$ than its physical velocity dispersion.

As a sanity check, we investigate the full distribution of our lens candidates' residual spectra $f_{\lambda,r}$, which were created by subtracting the REDROCK best-fit model from the observed DESI spectrum. In Figure 5, we divide the residual spectra $f_{\lambda,r}$ into 12 bins of 0.1 in $z_{\rm BG}$. In each one, we plot the median residual spectrum in the rest frame of the pipeline-identified background [O II] doublet. Given that the $z_{\rm BG}$ distribution is not uniform between 0.33 and 1.50 (Figure 4), separating the spectra into equal-width bins in $z_{\rm BG}$ space means higher $z_{\rm BG}$ bins include significantly more spectra than lower $z_{\rm BG}$ bins. Consequently, lower $z_{\rm BG}$ bins' median residual spectra are much noisier than those of higher $z_{\rm BG}$ bins.

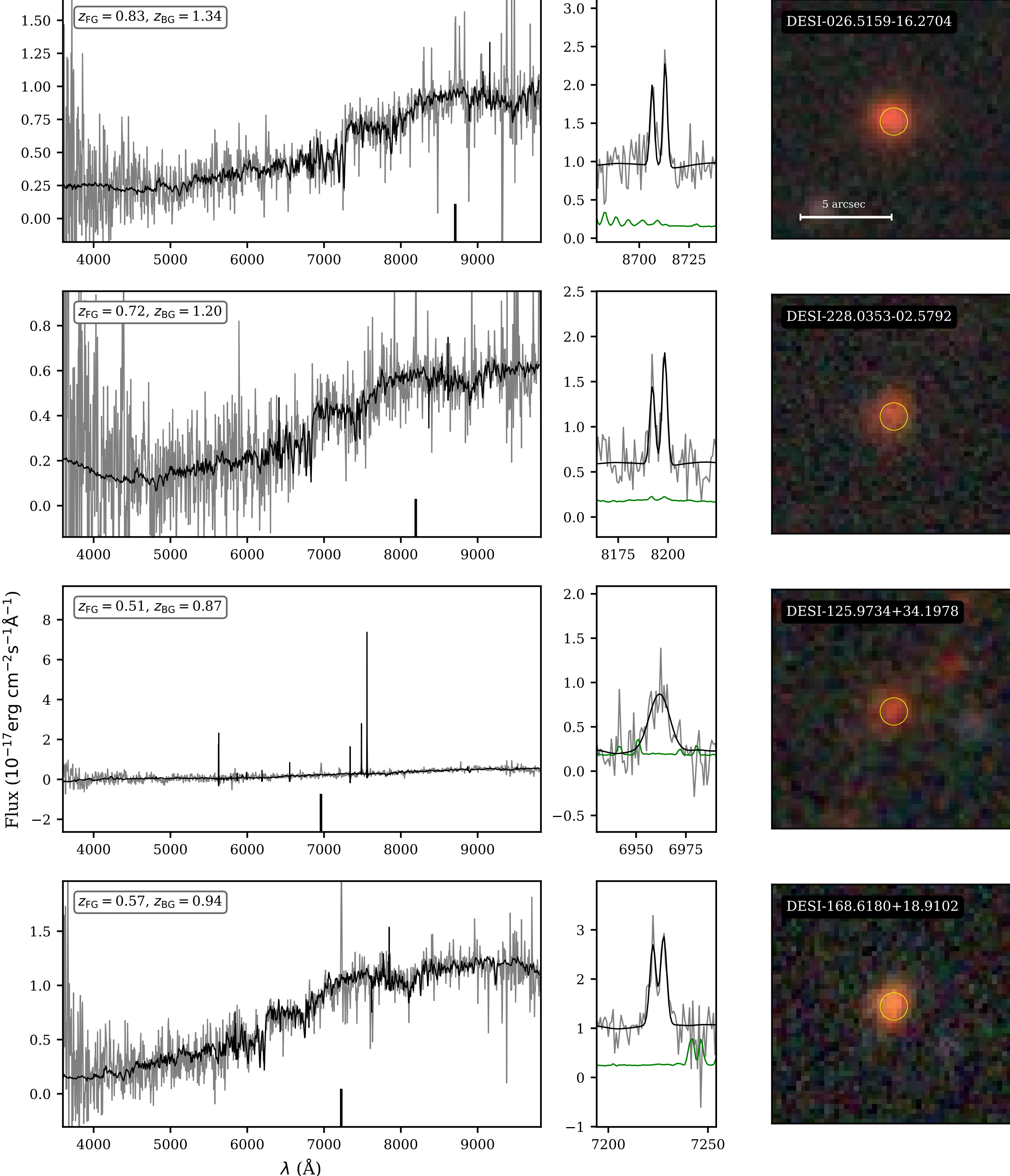

**Figure 3.** Panel 1/5. Examples of lens candidates, selected at random from among the full sample of 4110 candidates. Left subpanel: the smoothed LRG spectrum (gray) and REDROCK best-fit model spectrum (black) are overplotted. The wavelength of the background [O II] $\lambda\lambda$3727 doublet identified in the spectrum is indicated by a vertical black line up from the *x*-axis. The foreground LRG redshift ($z_{\rm FG}$) and background [O II] doublet redshift ($z_{\rm BG}$) are shown in the upper-left corner. Center subpanel: zoomed-in region on the background [O II] doublet identified in the LRG spectrum (gray), with our best-fit MCMC result (black) and REDROCK best-fit errors (green). Right subpanel: cutout from the DESI Legacy Imaging Surveys Data Release 10 (DR10), of size 13.″1 × 13.″1, where north is up and east is left. The DESI fiber size and placement are shown in yellow. The candidate's name (taken from its R.A. and decl. in degrees) is indicated in the upper-left corner. Panel 2/5. Examples of lens candidates. Full spectra (left), zoomed-in region on the background [O II] doublet (center), and cutouts from the DESI Legacy Imaging Surveys (right). Panel 3/5. Examples of lens candidates. Full spectra (left), zoomed-in region on the background [O II] doublet (center), and cutouts from the DESI Legacy Imaging Surveys (right). Panel 4/5. Examples of lens candidates. Full spectra (left), zoomed-in region on the background [O II] doublet (center), and cutouts from the DESI Legacy Imaging Surveys (right). Panel 5/5. Examples of lens candidates. Full spectra (left), zoomed-in region on the background [O II] doublet (center), and cutouts from the DESI Legacy Imaging Surveys (right).

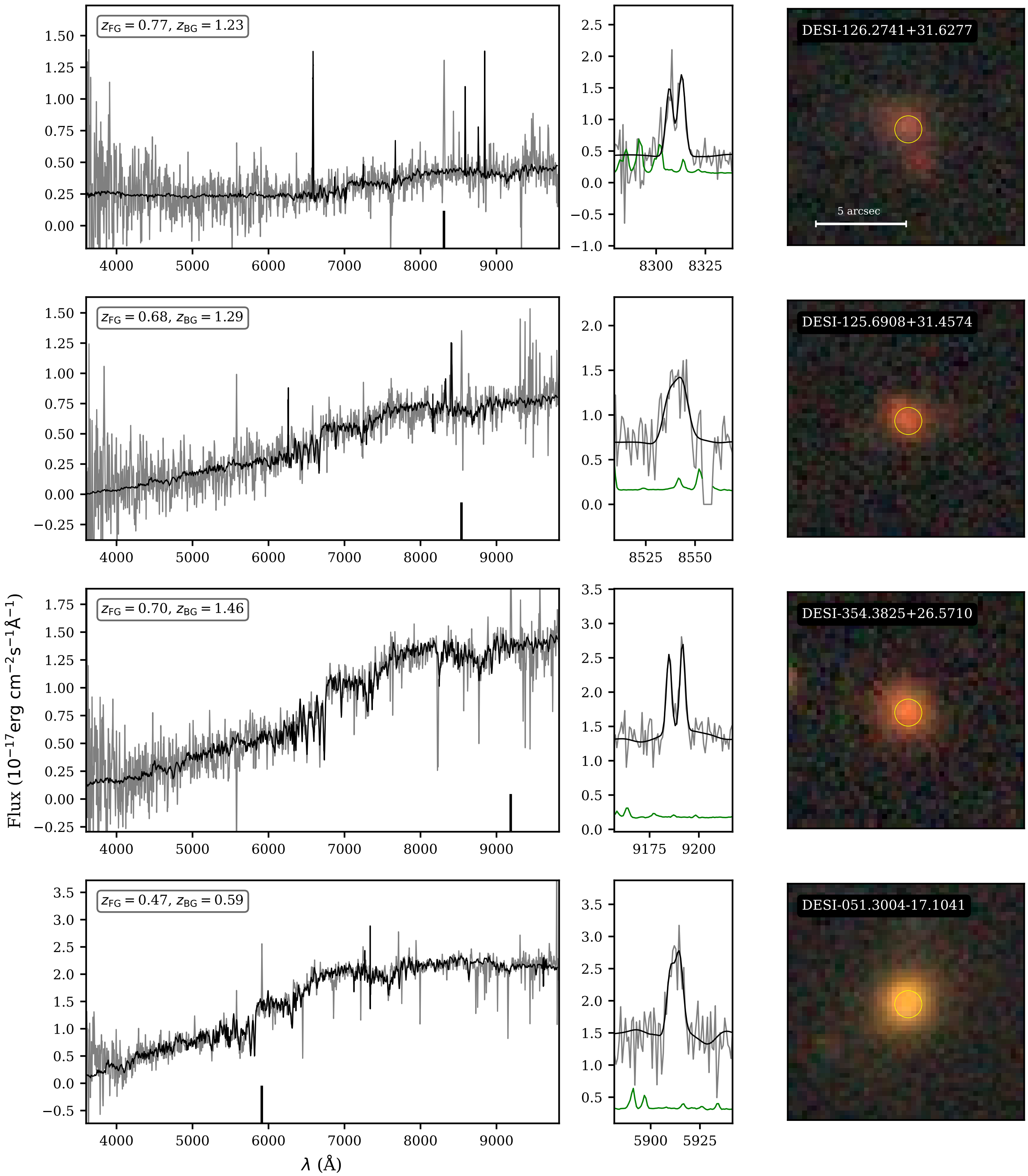

**Figure 3.** (Continued.)

In every redshift bin, the [O II] doublet stands out at 3727 Å as the strongest emission feature in the median residual spectrum, as we have specifically selected for it in our search. Other background emission lines are also visible in some redshift bins. In particular, we detect strong H$\gamma$ $\lambda$4340, H$\beta$ $\lambda$4861, [O III] $\lambda$4959, and [O III] $\lambda$5007 emission lines in many bins. We also detect weaker [Ne III] $\lambda$3869, [Fe V] $\lambda$3891, H$\delta$ $\lambda$4101, and potentially [Ne III] $\lambda$3967 in certain bins. An additional feature that is visible in some, but not all, bins is the 4000 Å break (G. Kauffmann et al. 2003). All of these additional background features are characteristic of star-forming galaxies at intermediate to high redshifts, the main sources whose light we expect would be lensed by foreground LRGs. The purity of our sample—that is, the percentage of our lens candidates whose pipeline-identified background feature is in fact a secondary [O II] $\lambda\lambda$3727 doublet from a source at a higher redshift than the foreground LRG, rather than a contaminant—is difficult to evaluate quantitatively. However, the presence of these additional emission features in the median residual spectra is evidence that the purity is quite high.

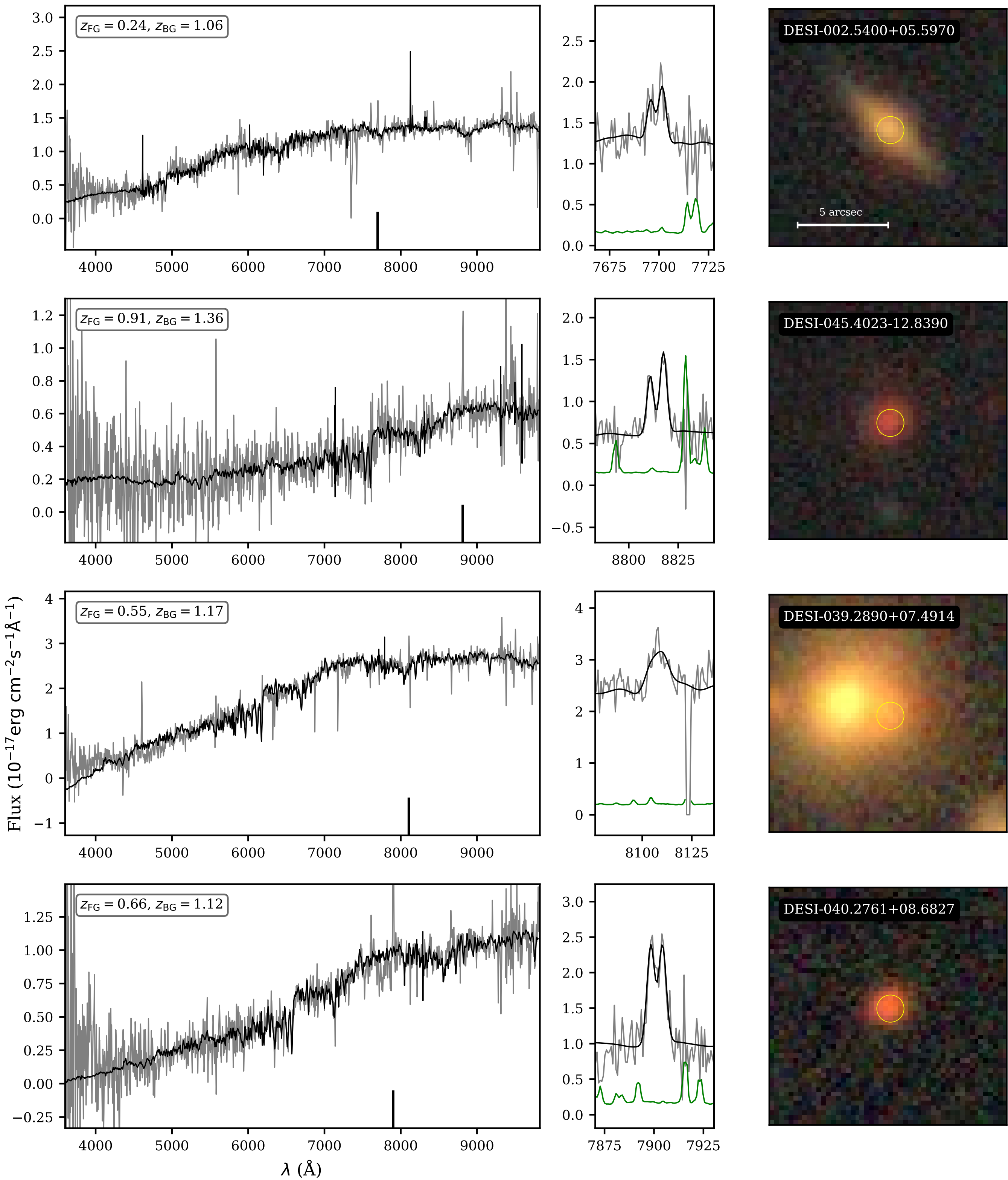

**Figure 3.** (Continued.)

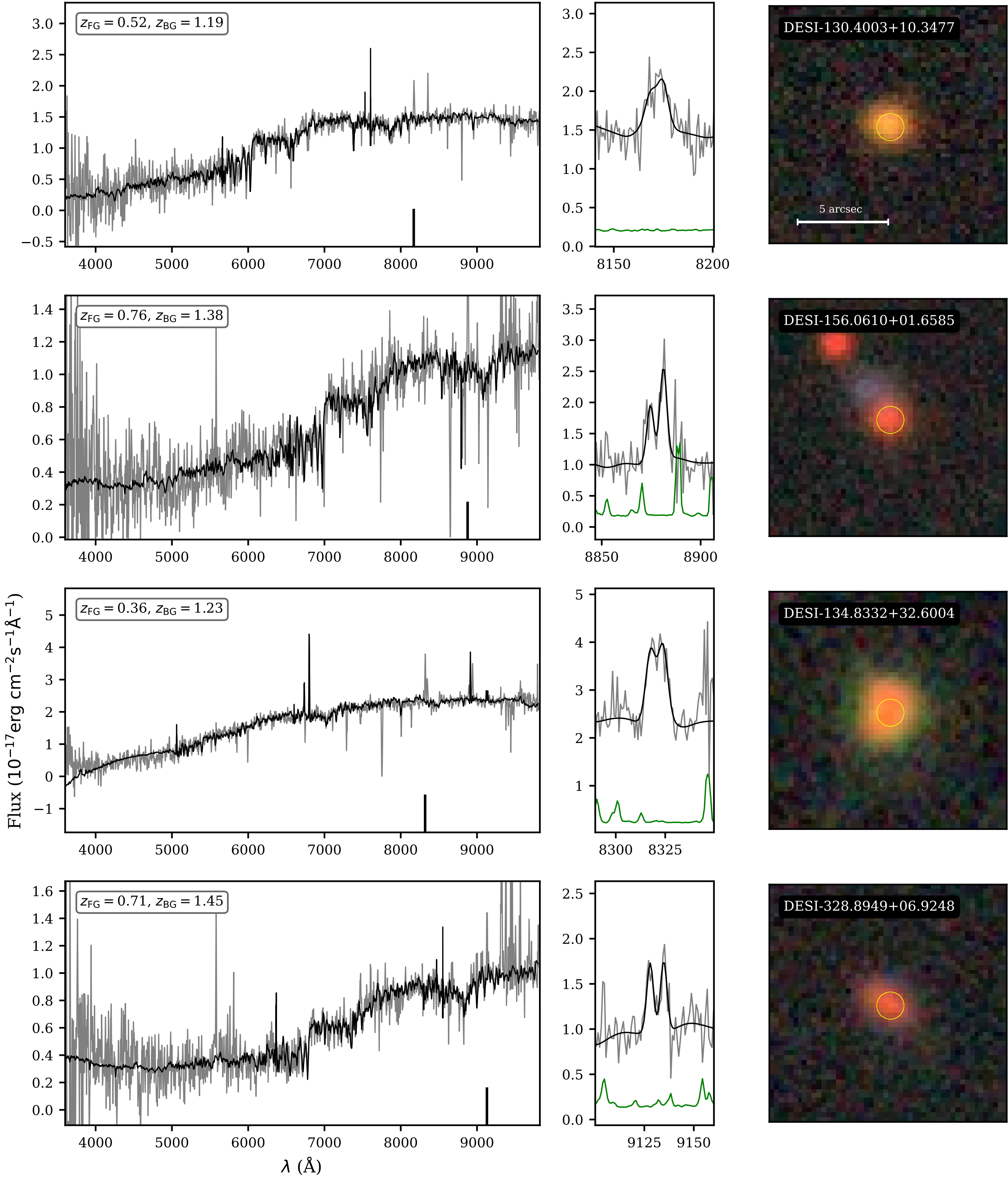

**Figure 3.** (Continued.)

In the right panel of Figure 5, we zoom in on the region around the [O II] $\lambda\lambda$3727 doublet. In every redshift bin, even the $z_{BG} < 0.4$ bin, which includes only $n = 4$ spectra, the double-peak shape of the emission feature is clearly visible. For all $z_{BG} > 0.6$ bins, the median 3726 Å line flux is consistently somewhat smaller than the median 3726 Å line flux, corresponding to a line amplitude ratio of roughly 1:1.3. Once again, this ratio aligns with previous empirical and physically motivated studies of the [O II] $\lambda\lambda3727$ flux ratio. It is worth noting that, as $z_{\rm BG}$ increases, the flux of the [O II] doublet and other background emission lines decreases. Thus, sources at higher background redshifts are apparently fainter, as expected. There also appears to be a slight anticorrelation between the redshift of the background source and the velocity dispersion of the detected [O II] doublet. This too is expected, since the velocity dispersion of a galaxy is directly related to its mass, such that more-massive galaxies have larger velocity dispersions. Given hierarchical structure formation, galaxies at higher redshifts are, on the whole, less massive than galaxies at lower redshifts.

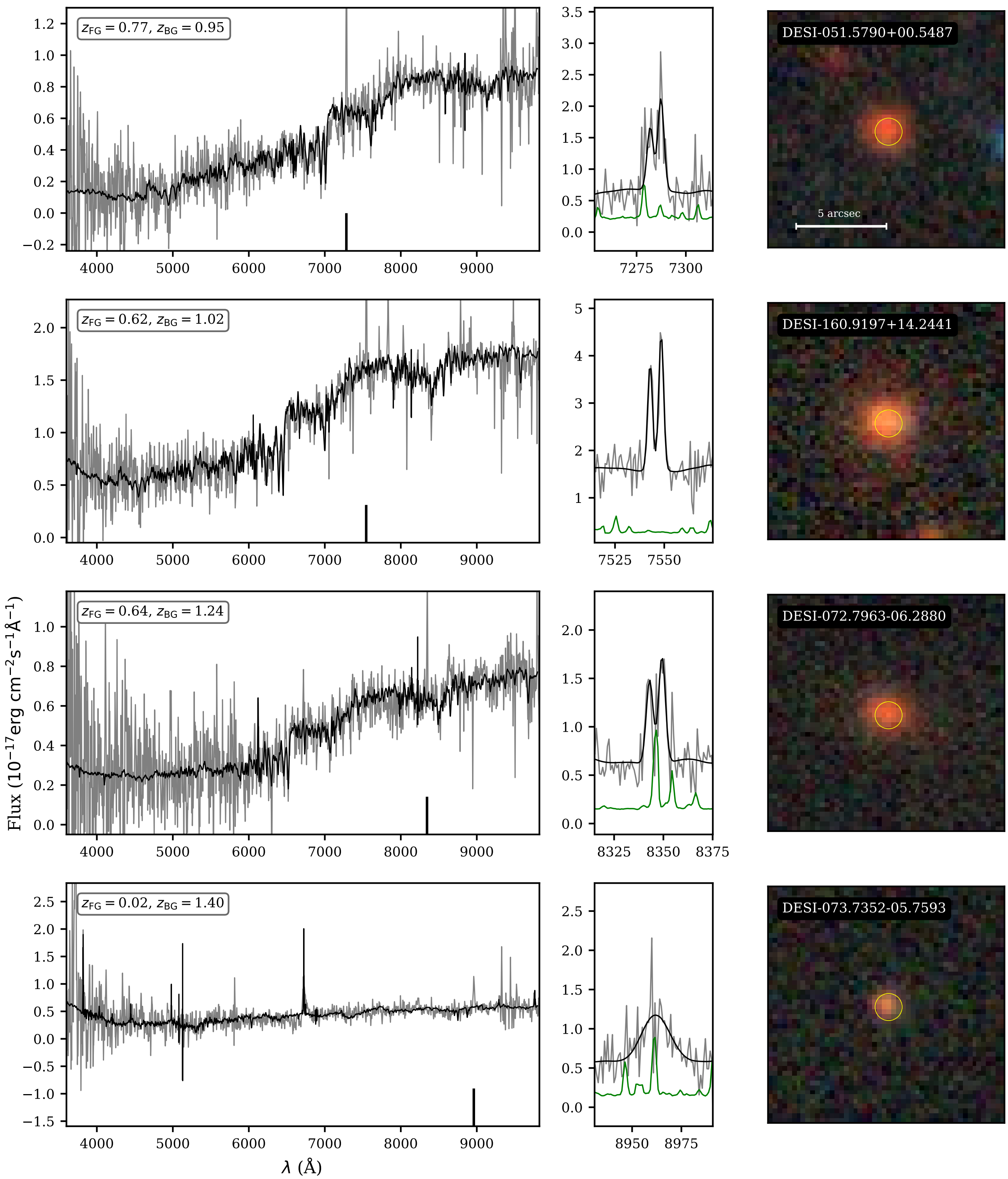

**Figure 3.** (Continued.)

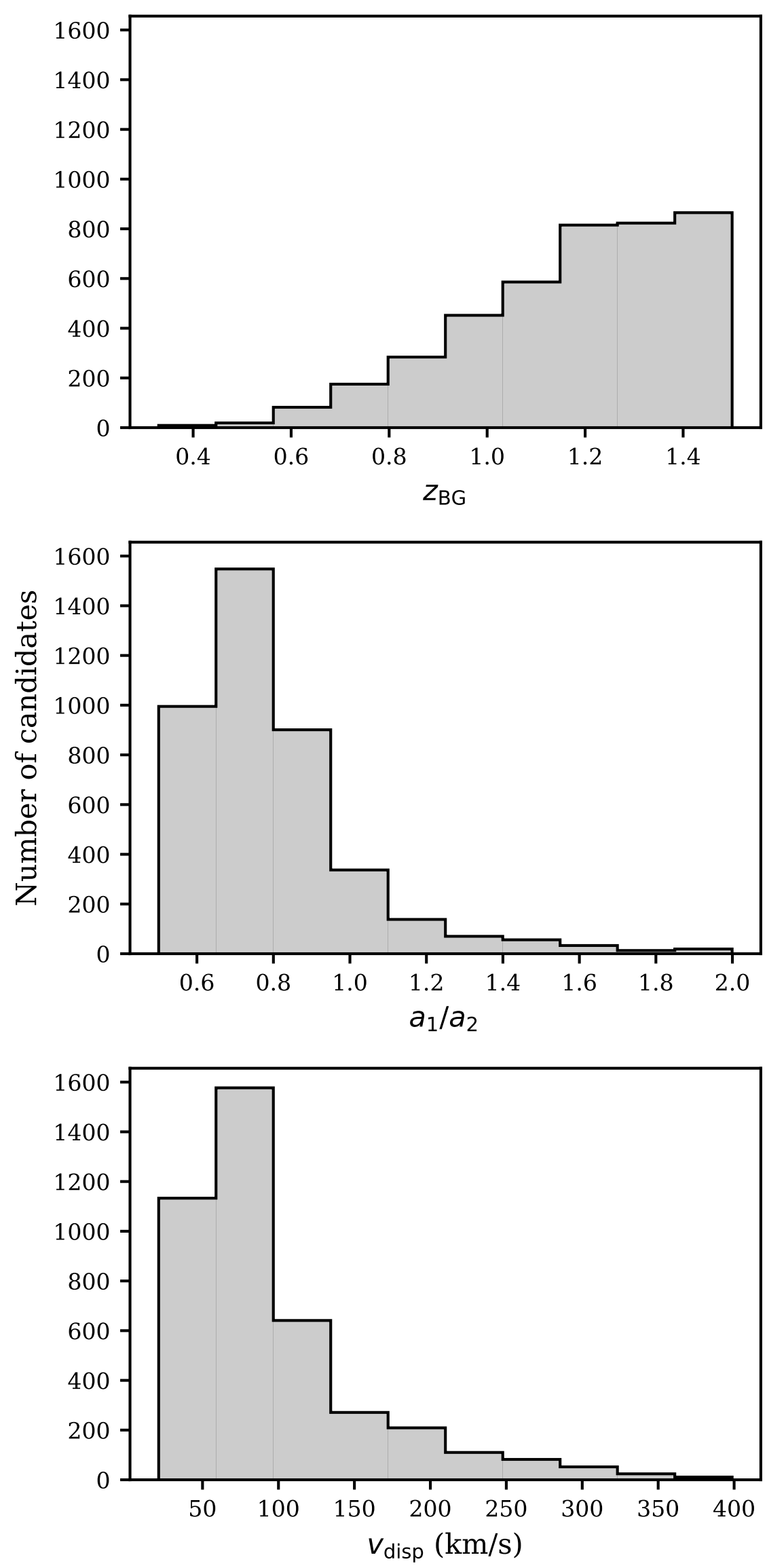

**Figure 4.** Distribution of recovered fit parameters for lens candidates' background [O II] $\lambda\lambda$3727. Top panel: redshift of the background doublet ($z_{BG}$), as identified by the initial linear fit. Middle panel: line flux ratio ($a_1/a_2$) between the two lines in the background doublet, as identified by the secondary MCMC fit. Bottom panel: velocity dispersion of the source ($v_{disp}$) corresponding to the width of the lines in the background doublet, as identified by the secondary MCMC fit.

## 5. Completeness

We quantify the completeness of our lens candidate sample as a function of background [O II] line flux, line width, and redshift. To do so, we inject simulated doublets into the DESI spectra of noncandidate LRGs, run our pipeline on these modified spectra, and measure the pipeline's recovery rate as a function of the injected parameters. We select the 5,786,191 DESI LRGs observed in *Loa* with good REDROCK coaddition fiberstatus ("COADD_FIBERSTATUS" = 0), which did not pass the three-phase selection pipeline. These form the baseline simulation sample. We then artificially inject fake background [O II] $\lambda\lambda$3727 doublets into each spectrum. We draw the parameters for the injected doublets from the following uniform distributions: line widths corresponding to velocity dispersions between 20 and 500 km s$^{-1}$, integrated [O II] $\lambda\lambda$3727 fluxes between 0.1 and 50 $\times$ 10$^{-17}$ erg s$^{-1}$ cm$^{-2}$, and background redshifts between the LRG's REDROCK-identified redshift $z_{FG}$ and 1.50. We assume a line flux ratio of 1:1.3. We then run the full three-phase candidate selection pipeline on these modified spectra. Because the unmodified spectra were selected to not present any features that make the pipeline cuts, any feature recovered at this stage must be the artificially injected doublet.

We report the results of our completeness simulations in Figure 6. Panels represent completeness as a function of injected integrated [O II] line flux and velocity dispersion, binned by background redshift. The completeness fraction depends minimally on $z_{BG}$, since noise associated with continuum flux varies little over DESI's observable wavelength range. The completeness fraction is strongly dependent on both the integrated line flux and velocity dispersion (i.e., the line width). The maximum completeness fraction of $\sim$0.7 is reached for [O II] doublets with injected velocity dispersions between 100 and 220 km s$^{-1}$ and integrated line fluxes between 25 and 35 $\times$ 10$^{-17}$ erg s$^{-1}$ cm$^{-2}$.

Above line widths of $\sim$100 km s$^{-1}$ at lower integrated line fluxes and $\sim$300 km s$^{-1}$ at higher integrated line fluxes, the completeness fraction drops off steeply. This is because the signal-to-noise threshold for detecting an emission feature is dependent on the peak spectral density line flux, not the integrated line flux. At fixed integrated line flux, doublets with higher velocity dispersions have lower peak line fluxes, since the total flux is smeared across a wider wavelength range. At lower integrated line fluxes, the minimum velocity dispersion threshold for this effect is lower, since a lower integrated flux spread over the same line width yields a lower spectral density line flux. Therefore, the maximum velocity dispersion detectability threshold evolves somewhat linearly as a function of integrated line flux.

At low velocity dispersions ($v_{disp} < 100$ km s$^{-1}$), the completeness fraction also falls off steeply. This is due to the hard cut we impose on the minimum velocity dispersion ($v_{disp} < 20$ km s$^{-1}$) to account for the instrumental resolution. This removes features that are nonphysical, i.e., that have line widths smaller than a delta function smeared by the instrumental resolution matrix. Features with such small widths are necessarily artifacts and are, therefore, excluded from the catalog.

At low ($<$15 $\times$ 10$^{-17}$ erg s$^{-1}$ cm$^{-2}$) integrated line fluxes, the completeness fraction drops significantly ($<$0.2), owing to the doublet's low signal-to-noise. It is worth noting that the majority of our lens candidates present [O II] doublets with integrated line fluxes below 10 $\times$ 10$^{-17}$ erg s$^{-1}$cm$^{-2}$. Therefore, if we are recovering $<$20% of real [O II] doublets at these line fluxes, we expect that the real number of background [O II] emitters behind DESI LRG fibers is 5 times greater than the number we discover.

Conversely, at high ($>$40 $\times$ 10$^{-17}$erg s$^{-1}$ cm$^{-2}$) integrated line fluxes, the completeness fraction also drops significantly. This is more puzzling, as one would assume that the completeness fraction would continue to increase with increased integrated line flux and, consequently, increased signal-to-noise ratio. Upon investigation, we find that this completeness drop-off at high line fluxes is in fact due to the hard selection cut of $\chi_r^2 < 3$. In our selection process, we fix $z_{BG}$ at the value recovered by the initial linear least-squares fit, and do not allow $z_{BG}$ jitter in the MCMC fitting phase. At higher integrated line fluxes, the same percentage offset between the data and the MCMC best-fit double-Gaussian model corresponds to a higher absolute offset, which then yields a higher $\chi_r^2$. In order to apply a truly consistent cut on goodness-of-fit across all doublets, we would need to either (a) introduce jitter in $z_{\mathrm{BG}}$ at the MCMC fitting phase, or (b) scale the maximum $\chi_r^2$ cut with recovered integrated line flux. However, neither of these effects change the lens catalog. To determine this, we visually inspect all 16 DESI LRG spectra with high integrated line fluxes that passed every cut except for $\chi_r^2 < 3$, and have $3 < \chi_r^2 < 20$. None of the background features identified in these spectra appear [O II] doublet-like. Therefore, we do not exclude any real candidates due to this completeness drop-off.

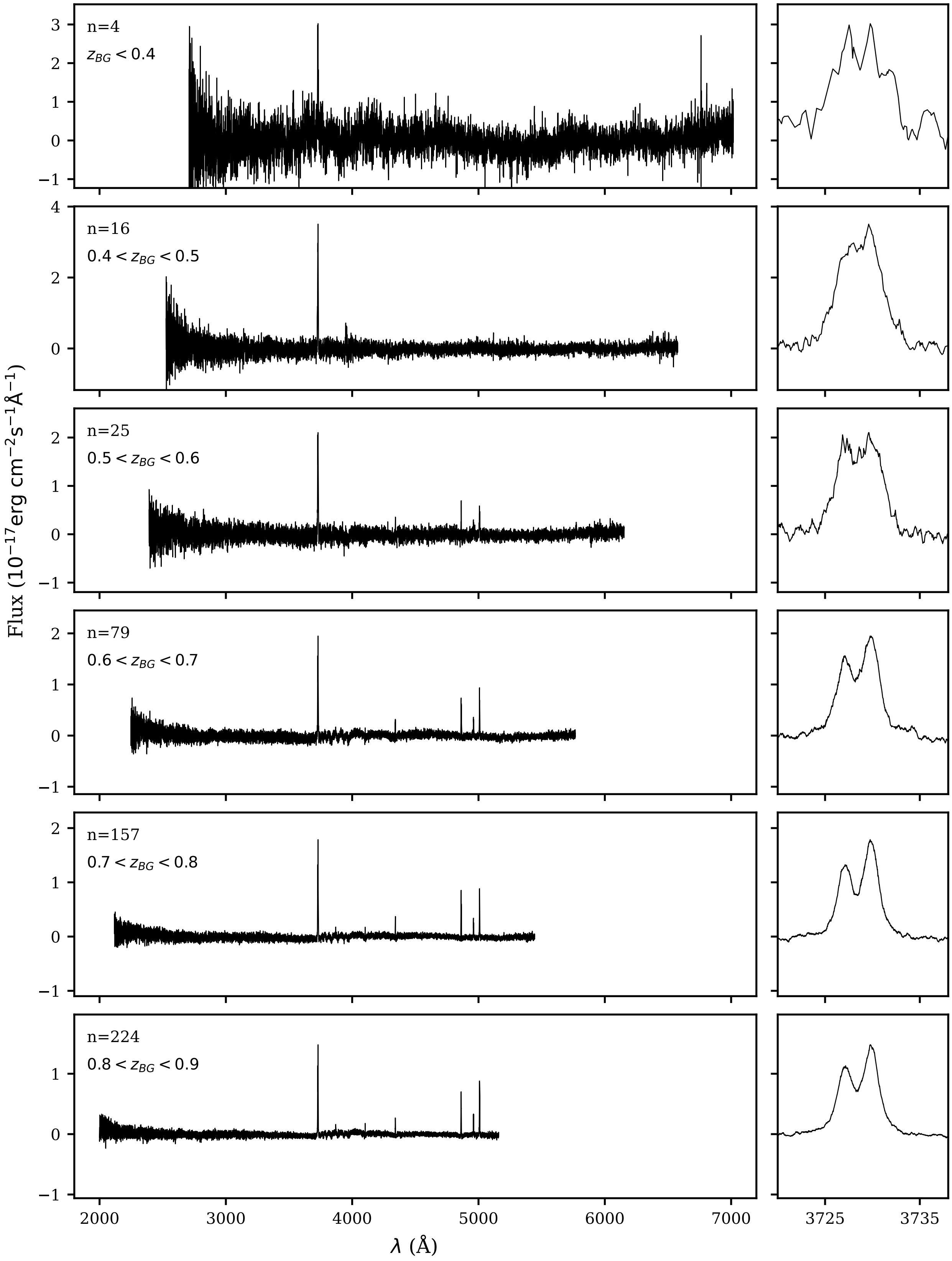

**Figure 5.** Panel 1/2. Median residual spectra, separated into bins of 0.1 in background redshift ($z_{\rm BG}$) space, of our lens candidates. Spectra are shown here in the rest frame of the background [O II] doublet identified by the selection pipeline. In each panel, the background redshift bin is indicated in the upper-left corner, alongside the number of spectra in this bin that were used to create the plotted medians ($n$). The leftmost panel of each row represents the full residual spectrum, while the rightmost panel shows a zoomed-in region on the [O II] $\lambda\lambda$3727 doublet. In bins with larger $n$, additional emission lines and the 4000 Å break are visible in the median residual spectra. Panel 2/2. Median residual spectra binned by background [O II] $\lambda\lambda$3727 doublet redshift ($z_{\rm BG}$), shown in the rest frame of the [O II] doublet. Left subpanel: full median residual spectra. Right subpanel: zoomed-in region on background [O II] doublet.

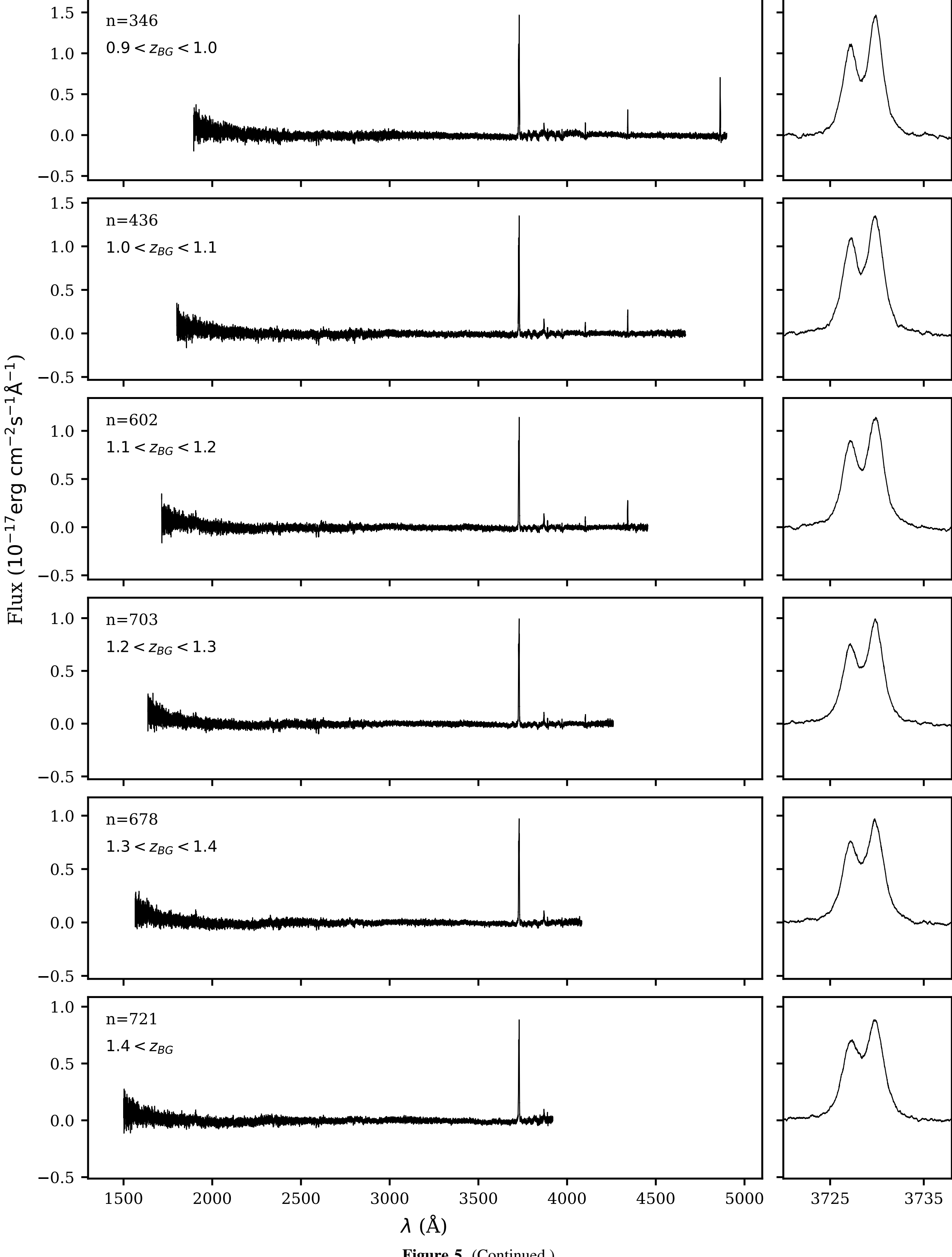

**Figure 5.** (Continued.)

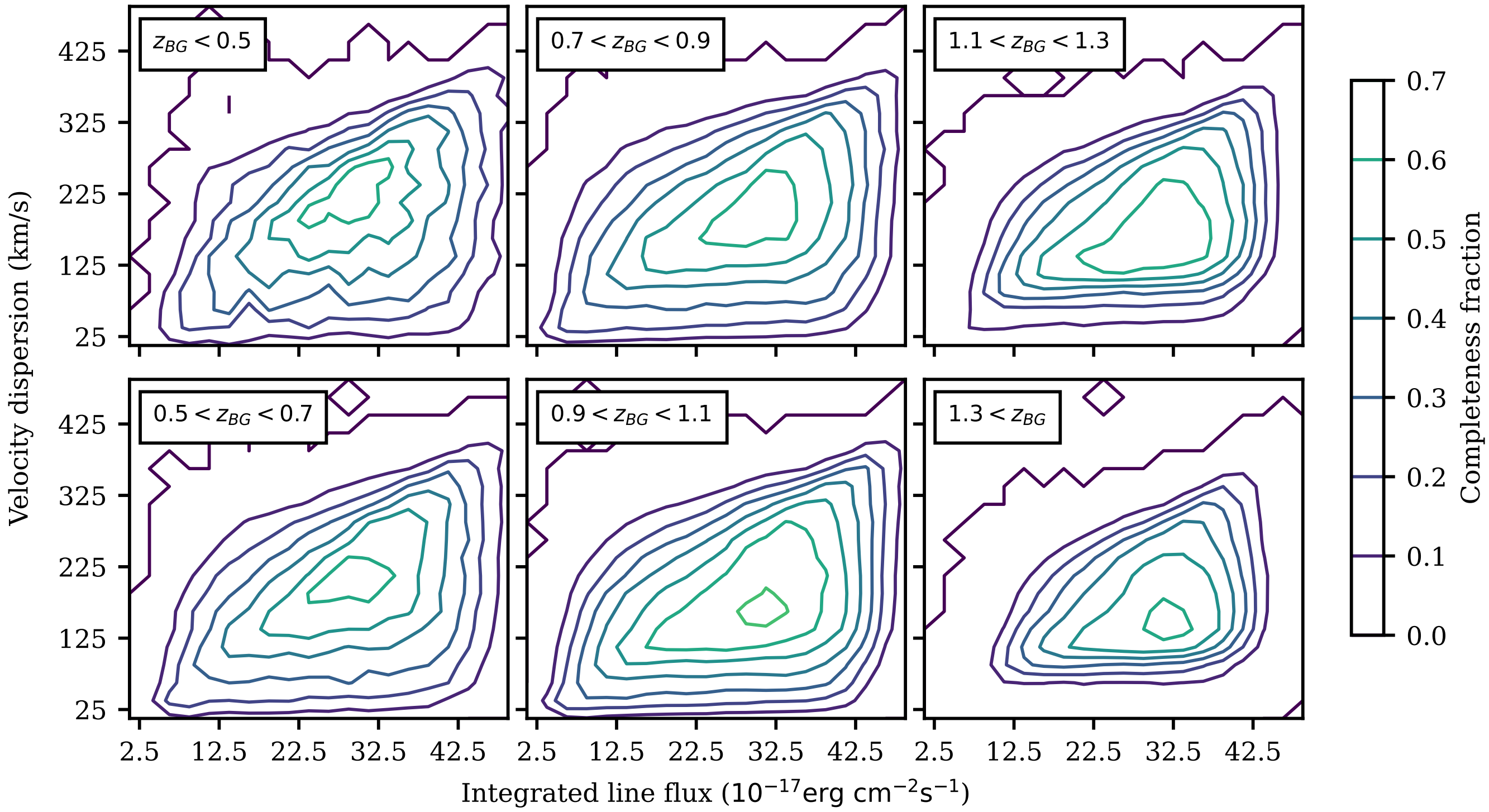

**Figure 6.** Completeness as a function of injected background [O II] doublet velocity dispersion and integrated line flux, binned by background redshift. The completeness fraction does not evolve much across background redshift bins. The maximum completeness of ∼0.7 is reached in each redshift bin for velocity dispersions between 100 and 220 km s$^{-1}$ and integrated line fluxes between 25 and 35 × $10^{-17}$ erg s$^{-1}$ cm$^{-2}$.

Across all bins, the completeness fraction reaches a maximum of only ∼0.7. Multiple effects contribute to the pipeline's inability to recover 100% of injected [O II] doublets. First, as explained in Section 3, to reduce the number of false-positive detections, we apply hard masks around strong emission lines in the foreground LRG spectrum. The sum of these masked wavelength ranges (see Table 1 for details) corresponds to, depending on the foreground LRG's redshift, as much as ∼6% of the wavelength range. We also apply hard masks around any areas of the DESI spectrum that have bad data points. The pipeline cannot recover doublets within these masks. These combined effects limit the maximum completeness to ∼0.9.

Another factor that limits the maximum completeness fraction is the $\Delta\chi_r^2 > 100$ cut we apply after the initial linear fit. This cut removes features that do not have significantly better reduced $\chi_r^2$ values than other portions of the residual spectrum. If the residual presents other background emission features in addition to the [O II] doublet, one of these features may be wide enough that the double-Gaussian model produces a decent fit. The doublet's $\Delta\chi_r^2$ would then not be large enough to pass the cuts. The sum of this effect and the hard masks are able to explain the completeness fraction's peak at ∼0.7.

## 6. Lenses or Not?

The sample is composed of 4110 LRGs selected to present [O II] emission from a secondary source at a higher redshift than the foreground LRG within the angular extent of a DESI fiber. We have detected emission from galaxies behind LRGs along the line of sight, but we have not yet determined whether the LRGs are acting as strong lenses to the background galaxies. In Section 6.1, we calculate the probability that each candidate is a lens. In Section 6.2, we discuss routes for lens confirmation via high-resolution imaging.

### 6.1. Lensing Probabilities

As an initial test of the ratio between real lensing systems and non-lensing galaxies superposed along the line of sight in our sample of lens candidates, we compare the distribution of velocity dispersions of our lens candidate LRGs to the distribution of LRG velocity dispersions in the full *Loa* catalog (Figure 7). To first order, a galaxy's velocity dispersion is a proxy for its mass. The Einstein radius is a function of foreground redshift, background redshift, and lens mass. At fixed distance, galaxies with higher masses are more likely to be lenses to background objects since their lensing cross sections are greater. Thus, we expect that the distribution of lensing LRGs may be skewed toward higher velocity dispersions than non-lensing LRGs (T. E. Collett 2015). DESI velocity dispersions are calculated by the FASTSPECFIT spectral fitting pipeline, the details of which are described in J. Moustakas et al. (2023). Convergent FASTSPECFIT velocity dispersion measurements exist for 2299 of our 4110 candidates (56%) and for 2,749,782 of the 5,837,154 LRGs in *Loa* (47%). For the remaining spectra with nonconvergent velocity dispersion measurements, the FASTSPECFIT pipeline assigns a default velocity dispersion value of 250 km s$^{-1}$, no longer treating it as a free parameter. Our candidates' velocity dispersion distribution is skewed slightly higher than that of the full DESI LRG sample, as expected for a sample of real lenses.

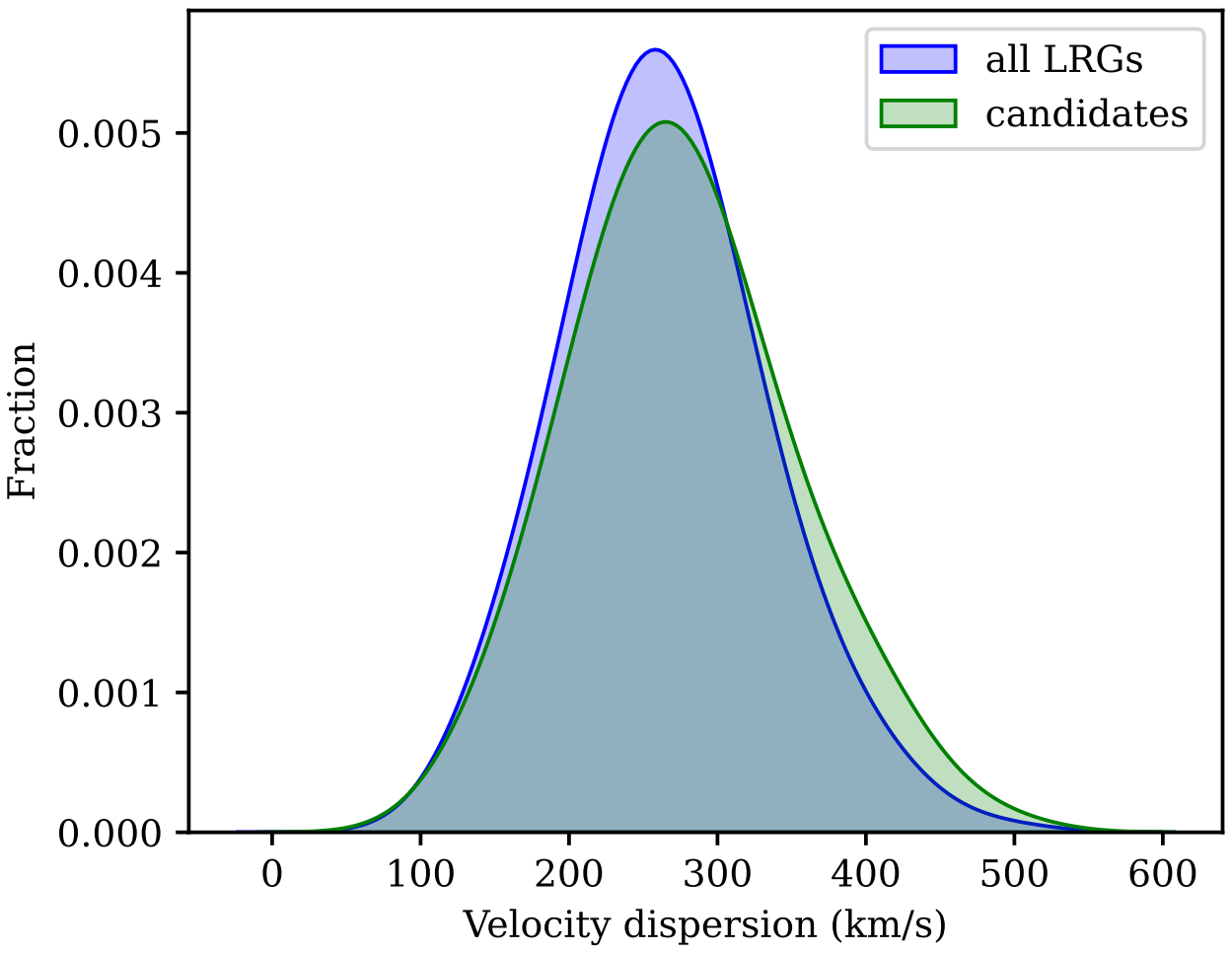

**Figure 7.** Comparison of FastSpecFit velocity dispersion distributions between the full *Loa* DESI LRG sample (blue) and LRGs included in our sample (green). Fractions are calculated using a Gaussian density kernel estimator with a bandwidth factor of 0.5. We only plot systems with convergent velocity dispersion measurements. The distribution of lens candidate LRGs' velocity dispersions is skewed to slightly larger values than that of all LRGs, indicating that our candidates may have preferentially larger masses, which we would expect if many of them are lenses.

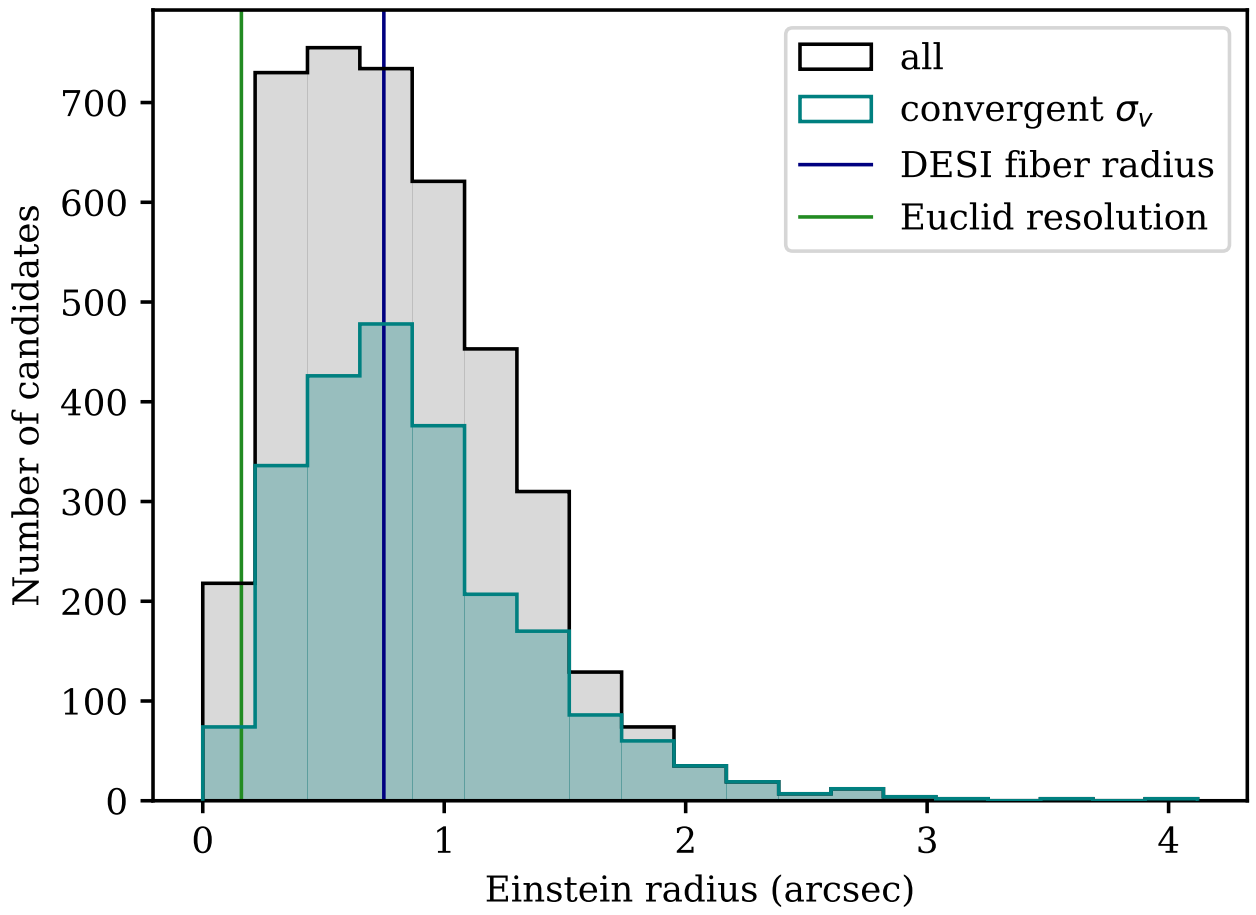

**Figure 8.** Distribution of all 4110 candidates' Einstein radii (gray), calculated from the foreground LRG redshift $z_{\rm FG}$ and background source [O II] doublet redshift $z_{\rm BG}$, and assuming a singular isothermal sphere model for the foreground LRG. The 2299 candidate systems with convergent FASTSPECFIT foreground velocity dispersion measurements are highlighted in teal. The radius of a DESI fiber (dark blue) and the Euclid space telescope's angular resolution (green) are overplotted.

Next, following similar methods to those described in A. S. Bolton et al.'s (2008) Appendix B, we more rigorously calculate the probability that each individual candidate is a lens rather than two non-lensing galaxies superposed along the line of sight. Whether a candidate system is a lens or not depends on whether the detected [O II] doublet is emitted by a source within the Einstein radius or outside it. We construct a probability density function (PDF) for the impact parameter $b$, or the position of the background source with respect to the foreground LRG, for each lens candidate. This PDF is a function of the observed foreground LRG redshift $z_{\rm FG}$, the background [O II] doublet redshift $z_{\rm BG}$, the velocity dispersion of the foreground LRG $\sigma_v$, the integrated line flux of the background [O II] doublet $S_{\rm obs}$, and the seeing of the DESI observation, as well as the cosmological redshift-dependent [O II] line luminosity function $\phi(L, z_{\rm BG})$.

We assume that the foreground LRG can be well modeled as a singular isothermal sphere. Under this assumption, as shown in R. Narayan & M. Bartelmann (1996), the Einstein radius $\theta_{\rm E}$ depends only on the angular diameter distance from the observer to the source ($D_{\rm s}$), the angular diameter distance from the lens to the source ($D_{\rm ls}$), and the velocity dispersion of the lens ($\sigma_v$).

$$\theta_{\rm E} = 4\pi \frac{\sigma_v}{c^2} \frac{D_{\rm ls}}{D_{\rm s}}. \tag{6}$$

Because our candidates are selected spectroscopically, we already have very accurate measurements of $z_{\rm FG}$ and $z_{\rm BG}$, as well as the velocity dispersion of the foreground LRG from the FASTSPECFIT DESI secondary fitting pipeline. We assume the Planck 2018 instance of a flat ΛCDM cosmological model (Planck Collaboration et al. 2020), and use ASTROPY's built-in cosmological functions (Astropy Collaboration et al. 2013, 2018, 2022) to calculate the angular diameter distances $D_{\rm s}$ and $D_{\rm ls}$ from $z_{\rm FG}$ and $z_{\rm BG}$. Using Equation 6, we calculate Einstein radii, and therefore lensing cross sections, for all 2299 candidate systems with convergent FASTSPECFIT velocity dispersions (Figure 8). We also calculate Einstein radii for systems with fixed, nonconvergent velocity dispersions of 250 km s$^{-1}$, but we warn that these values are much less reliable than those calculated from convergent FASTSPECFIT velocity dispersion results. As demonstrated in A. S. Bolton et al. (2008), lensing cross sections calculated in this manner correlate strongly with the eventual lens confirmation rate.

Next, we construct a redshift-dependent cosmological [O II] line luminosity function. We assume that this luminosity function can be described by a Schechter function (P. Schechter 1976), given here in log form:

$$\phi(L)\, d\log(L) = \phi^* \ln(10) \left(\frac{L}{L^*}\right)^{\alpha+1} e^{-(L/L^*)}\, d\log(L) \tag{7}$$

where the free parameters $\phi^*$, $L^*$, and $\alpha$ are, respectively, the characteristic density, luminosity, and gradient of the faint-end slope of the function. We make use of A. B. Drake et al.'s (2013) measurements of these parameter values for the [O II] line luminosity function at five discrete redshifts spanning from $z = 0.35$ to $z = 1.64$, which they report in their Table 4 and plot in their Figure 9. A. B. Drake et al. (2013) constructed their luminosity functions from narrowband-selected emission-line galaxies in the Subaru/XMM-Newton Deep Field (H. Furusawa et al. 2008) and the Ultra-Deep Survey (O. Almaini et al. 2007; S. Foucaud et al. 2007) field of the UK Infrared Telescope's Infrared Deep Sky Survey (UKIDSS; A. Lawrence et al. 2007). The $z = 0.53$ luminosity function suffers from low number statistics due to the shallowness of the NB570 Subaru filter, which is used at this redshift, causing its shape to be different from the other redshifts' luminosity functions. Using their reported volume elements and the Planck 2018 instance of ASTROPY's cosmology module, we translate their reported [O II] luminosity functions from units of volume density (Mpc$^{-3}$dex$^{-}$1) versus luminosity (W) to units of on-sky number density (deg$^{-2}$ dex$^{-1}$) versus integrated line flux ($10^{-17}$ erg s$^{-1}$ cm$^{-2}$), which we plot in Figure 9. For reference, we overplot DESI's reported integrated line flux limit ($8 \times 10^{-17}$ erg s$^{-1}$ cm$^{-2}$) and the lower limit of our candidates' detected [O II] doublets ($1.5 \times 10^{-17}$ erg s$^{-1}$ cm$^{-2}$). For each system, we linearly interpolate the reported values of $\phi^*$, $L^*$, and $\alpha$ to create a corresponding [O II] line luminosity Schechter function at any redshift between $z = 0.35$ and $z = 1.64$.

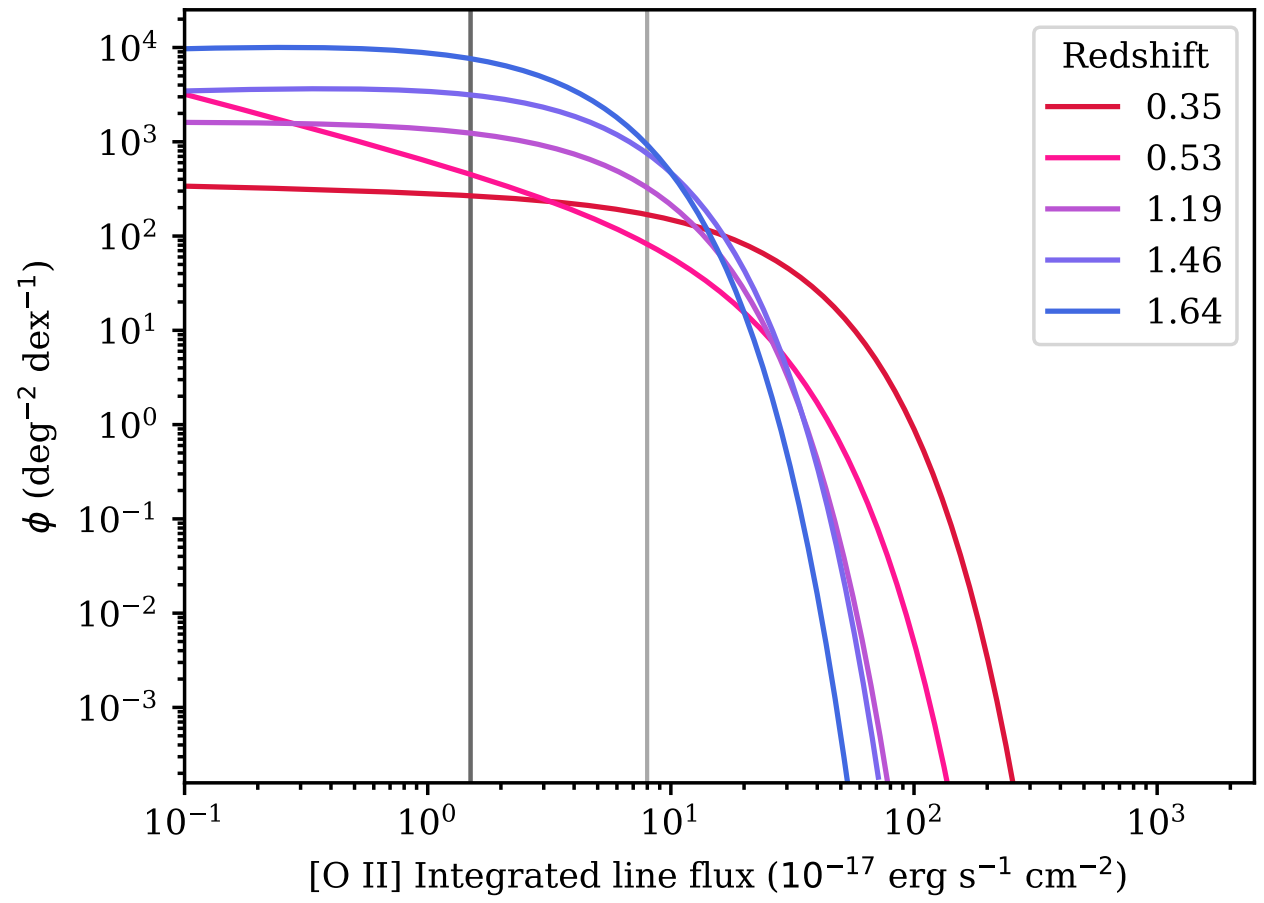

**Figure 9.** Redshift-dependent [O II] luminosity function (number density of [O II] emitters per on-sky area as a function of the doublet's integrated line flux). These Schechter function fits are calculated by A. B. Drake et al. (2013) using data from the UKIDSS Ultra Deep Survey. We overplot DESI's reported redshift success line flux limit ($8 \times 10^{-17}$ erg s$^{-1}$ cm$^{-2}$, light-gray vertical line), and the line flux limit down to which we are able to consistently identify background [O II] doublets in our search ($1.5 \times 10^{-17}$ erg s$^{-1}$ cm$^{-2}$, dark-gray vertical line).

For each candidate system, we use the LENSTRONOMY Python package (S. Birrer & A. Amara 2018; S. Birrer et al. 2021) to simulate the on-sky flux distribution $S_{\rm flux}$ from a lensed background source. We assume the foreground lens is well modeled by a singular isothermal sphere with the Einstein radius calculated as described above, and integrate across the impact parameter $b$ from $0'' < b < 3''$. To construct the point-spread function (PSF), we use DESI's recorded seeing for each observation. When this is not available, we assume an FWHM PSF value of $1\farcs1$, DESI's average seeing. For each system, we also simulate the flux from an unlensed background source $S$ across the same range of impact parameters by setting the Einstein radius to $0''$. We assume the background source has a half-light radius of $0\farcs35$, which is standard for emission-line galaxies observed by DESI at the same redshifts as our background sources. We compare the flux as a function of impact parameter $b$ in the lensed simulation instance and the unlensed instance. This gives us the relationship $f(b)$ between the lensed flux $S_{\rm obs}$ and intrinsic flux $S$ through the fiber as a function of impact parameter. Following Appendix B of A. S. Bolton et al. (2004), we combine $f(b)$ and the redshift-dependent [O II] line luminosity function $\phi$ to construct the joint PDF:

$$p(b;\, z_{\rm BG},\, S_{\rm obs})\, db = N(z_{\rm BG},\, S_{\rm obs})\, \frac{b}{f(b)} \cdot \phi\left(z_{\rm BG},\, \frac{S_{\rm obs}}{f(b)}\right) db \qquad (8)$$

where $N(z_{\rm BG},\, S_{\rm obs})$ is a normalization factor whose value proves unimportant.

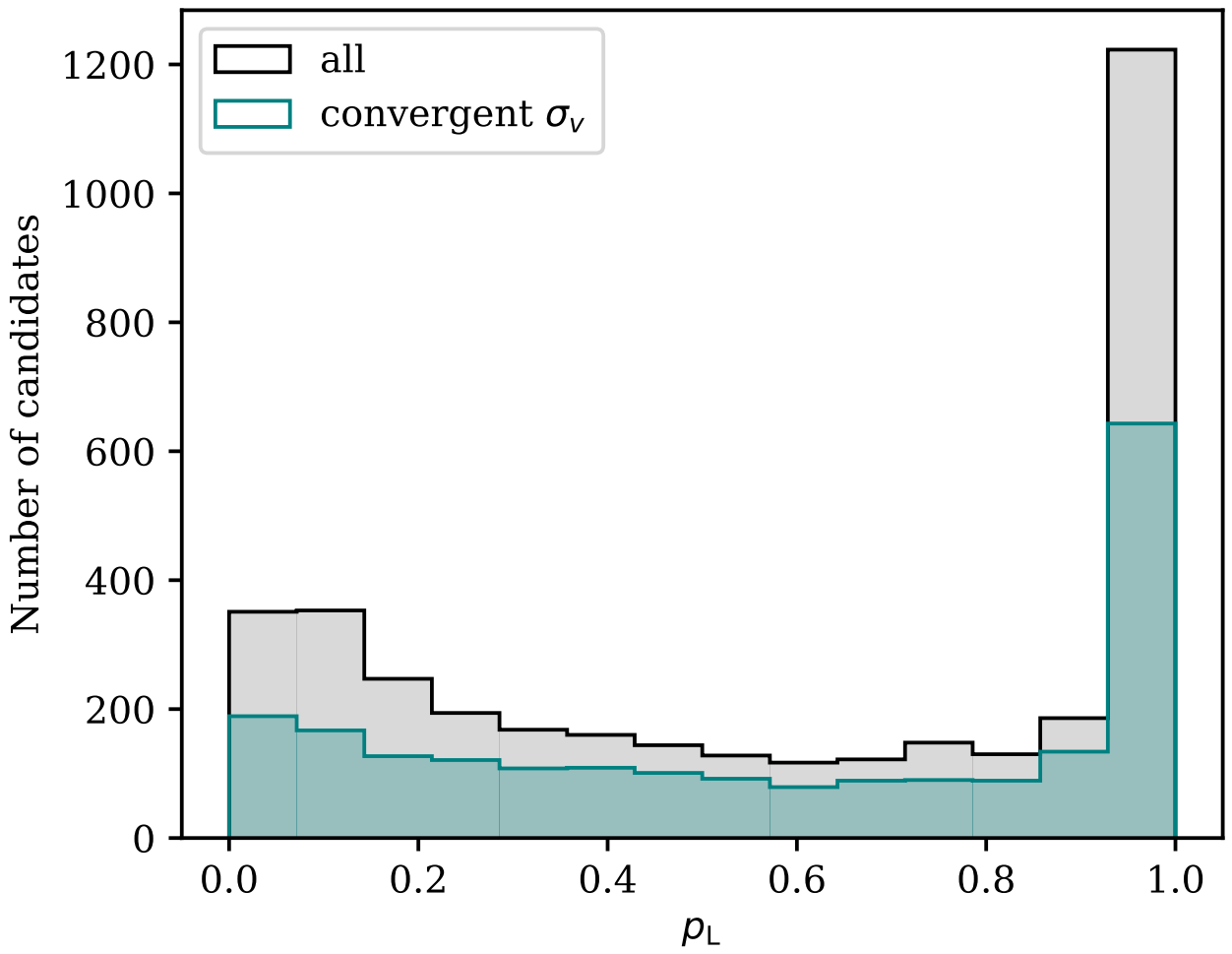

**Figure 10.** Distribution of lensing probabilities for all 4110 lens candidate systems (gray), with the subsection of 2299 candidate systems with convergent foreground LRG velocity dispersion measurements overplotted (teal). Probabilities are calculated according to Equation 9. Summing over these probabilities, we expect 2165 candidates (∼53%) to be lenses.

In order to calculate the probability that each candidate is a lens, we integrate $p(b;\, z_{\rm BG},\, S_{\rm obs})\, db$ over $0 < b < \theta_{\rm E}$. This yields the normalized probability $p_{\rm E}$ that a lensed background source falls within the lensing cross section, i.e., the on-sky disk with the system's Einstein radius centered on the foreground LRG. We then integrate $p(b;\, z_{\rm BG},\, S_{\rm obs})\, db$ over $\theta_{\rm E} < b < 3''$ to calculate the probability $p_{\rm A}$ that a background source of the measured redshift and integrated line flux falls within the annulus outside the lensing cross section. The probability $p_{\rm L}$ that the system is a lens rather than the chance alignment of two non-lensing galaxies along the line of sight is then

$$p_{\rm L} = \frac{p_{\rm E}}{p_{\rm E} + p_{\rm A}}. \qquad (9)$$

The normalization factor drops out in the division. For each system, $p_{\rm L}$ assumes a value between 0 and 1. We report the distribution of our candidates' lensing probabilities in Figure 10. While our candidates' lensing probabilities span the full range from 0–1, the distribution is weighted toward higher $p_{\rm L}$. Summing over all $p_{\rm L}$, we expect 2165 (∼53%) of our candidates to be real lenses. Note that 74 candidates have lensing probabilities of 1, and 1311 candidates have $p_{\rm L} > 0.9$.

As a final test, we limit the sample to the 3671 candidate systems with $z_{\rm BG} > 1.3\, z_{\rm FG}$ and $\theta_{\rm E} > 0\farcs1$, cuts that we expect most observable lenses to obey. The Euclid telescope is finding lensing systems with $\theta_{\rm E} \sim 0\farcs2$, close to its resolution limit of $0\farcs16$ (Euclid Collaboration et al. 2025b), and the smallest lens observed with the Hubble Space Telescope (HST) thus far has an Einstein radius of $\theta_{\rm E} = 0\farcs17$ (A. Barnacka et al. 2016; A. Goobar et al. 2023; J. D. R. Pierel et al. 2023). We predict that 2142 (or 58%) of these systems are real lenses, and should be confirmable with high-resolution imaging.

### 6.2. High-resolution Imaging Confirmation

Fully confirming spectroscopically identified lensing systems requires follow-up higher-resolution imaging. The instrument must have strong enough resolving power and low enough seeing to probe subarcsecond features, as a DESI fiber spans only $1\farcs5$ in diameter. As explained in Section 4, DESI LS imaging in many cases lacks sufficient angular resolution to detect the lensed source. Small numbers of spectroscopically identified lens candidates have been confirmed using integral field spectroscopy units such as those installed on the Gemini Observatory's Gemini Multi-Object Spectrographs (I. M. Hook et al. 2004) and high-resolution narrowband imaging with space-based instruments such as HST's Advanced Camera for Surveys (ACS; e.g., A. S. Bolton et al. 2006; J. R. Brownstein et al. 2012). However, no archival HST data exists for any of our candidates,[50] and these confirmation methods are not well suited for sample sizes as large as 4110 systems.

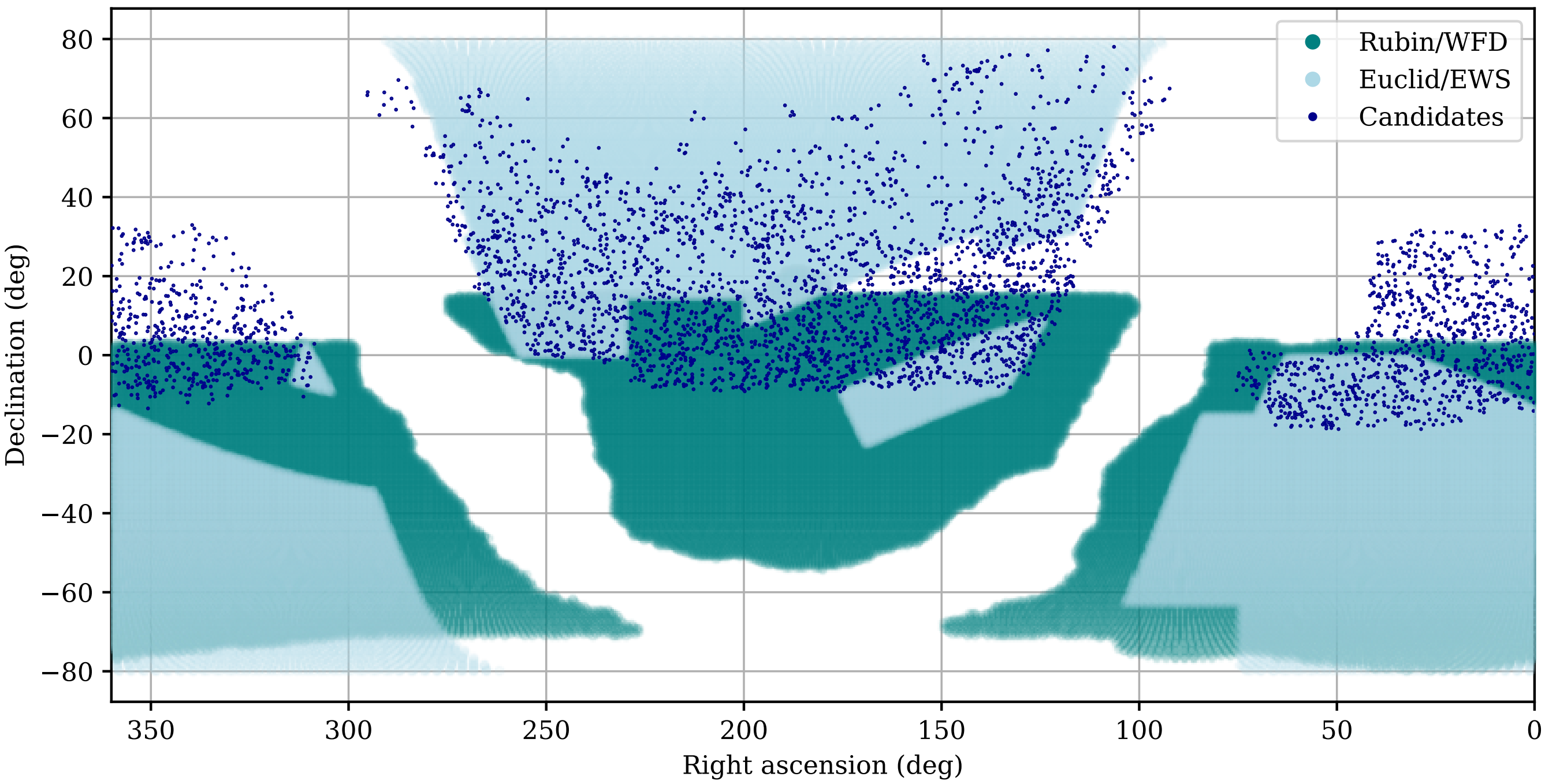

**Figure 11.** On-sky positions of all 4110 lens candidates (dark blue) over the planned Euclid Wide Survey (Euclid/EWS, light blue) and Vera Rubin Observatory's Wide Fast Deep survey (Rubin/WFD, teal) footprints. These surveys will likely have sufficient depth and resolution to confirm lens candidates they observe.

We turn instead to current and upcoming high-resolution wide-field imaging surveys. Over 10 yr, the Vera C. Rubin Observatory will conduct the LSST (Ž. Ivezić et al. 2019), observing 18,000 $deg^2$ uniformly every ∼3 nights for a total of 800 visits per position on the sky. The final coadded Rubin map will go to a depth of $r \sim 27.5$, spanning six optical filters (*ugrizy*). Rubin's median seeing will be $0\farcs7$. Rubin's main survey operations are expected to begin soon. As shown in Figure 11, roughly 2000 of our lens candidates fall within the anticipated Rubin/WFD footprint (teal), and many will therefore be able to be confirmed via Rubin imaging.

Similarly, the Euclid telescope's (R. Laureijs et al. 2011; Euclid Collaboration et al. 2025a) Euclid Wide Survey (EWS; Euclid Collaboration et al. 2022) is observing 15,000 $deg^2$ over the course of 6 yr down to a depth of $i \sim 26.2$. Euclid has the distinct advantage of being a space-based observatory, meaning that its angular resolving power is not limited by atmospheric seeing. The first Euclid Quick Data Release (Q1; Euclid Collaboration et al. 2025c) is already public. Six of our 4110 candidates are included in Q1 (Figure 12). Two of these six (panels (b) and (f)) present obvious lensed arcs (denoted by green arrows) and are therefore confirmed lensing systems. Two others (panels (d) and (e)) are edge-on and present a few small specks, which could be indicative of lensing, but require higher signal-to-noise and/or additional color imaging to confirm them. One (panel (a)) does not have high enough signal-to-noise, and one (panel (c)) is promising, but the nearby bright star complicates the confirmation. As shown in Figure 11, roughly 1300 of our candidates fall within the eventual Euclid/EWS footprint (light blue), and can be confirmed in this manner. Candidates that fall within the footprints of both Rubin/WFD and Euclid/EWS are particularly useful, as they can be used to compare the effectiveness of Rubin and Euclid imaging for the confirmation of future strong lenses.

The Nancy Grace Roman Space Telescope (D. Spergel et al. 2015) will also conduct a high-resolution, wide-field sky survey in the infrared, and can be used to confirm any candidates that fall within its footprint. Its survey area is not yet finalized.

## 7. Discussion

To better qualify our candidate sample, we compare our lens candidates to existing strong lens catalogs. We cross-match our candidates with a very nearly complete list of 12,210 to-date known lensing systems, including both lenses identified in imaging and spectroscopic surveys. We find that 223 of our 4110 candidates have previously been identified as lenses. The other 3887 are new discoveries. Some of these 223 overlapping systems have been previously identified in more than one lens search. Throughout this section, we consider sources to be the same object if their angular separation is $<3''$.

### 7.1. Overlap with Imaging Surveys

The DESI Lens Foundry has been using neural networks to identify lenses in the ground-based DESI LS. X. Huang et al. (2020; Paper I of the DESI Lens Foundry) discovered 335 candidate lenses in the Data Release 7 of DECaLS, a component of DESI LS. They employed a deep residual neural network, building upon work by F. Lanusse et al. (2018), to search for lens-like features, which they followed up with visual inspection. Only one of our 4110 lens candidates overlaps with the X. Huang et al. (2020) sample.

[50] https://mast.stsci.edu/search/ui/#/hst

X. Huang et al. (2021; Paper II of the DESI Lens Foundry) expanded the same search methods to the DECaLS Data Release 8 to identify 1210 additional lens candidates. Five of these are included among our 4110 candidates. C. Storfer et al. (2024, Paper III of the DESI Lens Foundry)[51] applied the neural network search to the DESI LS Data Release 9, covering 19,000 $\deg^2$ of the Northern Hemisphere sky. They identified 1895 lens candidates, of which 1512 were new discoveries. Eleven of their sources overlap with our lens candidates.

J. C. Inchausti et al. (2025, Paper IV of the DESI Lens Foundry)[52] expanded to the Southern Hemisphere, searching the 14,000 $\deg^2$ DESI LS DR10. Employing both a residual neural network and an EfficientNet model, they reported 811 new lens candidates. Only one of these overlaps with our sample. In total, 18—or just 0.4%—of our candidates overlap with those discovered in DESI LS imaging by the DESI Lens Foundry. Our single-fiber spectroscopic search systematically selects lenses with Einstein radii at or below the DESI fiber radius ($0''.75$), which is also below the DESI LS typical PSF of $0''.1$ (A. Dey et al. 2019). Therefore, the vast majority of our candidates are unresolved in the DESI LS (which is apparent in Figure 3).

Gravitational Lenses in UNIONS and Euclid (GLUE; C. J. Storfer et al. 2025) employed a similar residual neural network architecture to search for strong lenses in the Ultraviolet Near-Infrared Optical Northern Survey (UNIONS; S. Gwyn et al. 2025). UNIONS includes multiband data from the Canada–France–Hawai'i Telescope's Canada France Imaging Survey, the Subaru Telescope's Waterloo-Hawai'i-IfA *G*-band Survey and its Wide Imaging with Subaru HSC of the Euclid Sky survey, and the Panoramic Survey Telescope And Rapid Response System. GLUE reports 1346 new lens candidates, 20 of which overlap with our search. Separately, a previous one-filter UNIONS lens search (E. Savary et al. 2022) found 104 new lens candidates. None of these overlap with our sample.

The Euclid mission is developing its own pipeline to automatically photometrically detect and model lensing systems (The Strong Lensing Discovery Engine, Euclid Collaboration et al. 2025d; data access: M. Walmsley et al. 2025). In Q1, they reported 497 strong lenses, 243 of which are new discoveries. Only one of these overlaps with our sample, despite the fact that we see obvious lensing features in two Euclid images of the six candidates that were observed in Q1 (Figure 12). The candidate that Euclid's Strong Lensing Discovery Engine also recovers is the candidate shown in panel (f). The candidate shown in panel (b), which also shows obvious lensing features, is not identified by Euclid's Strong Lensing Discovery Engine.

[51] Data access: https://sites.google.com/usfca.edu/neuralens/publications/lens-candidates-storfer-et-al-2022?authuser=0

[52] Data access: https://sites.google.com/usfca.edu/neuralens/publications/lens-candidates-inchausti-et-al-2025?authuser=0.

### *7.2. Overlap with Spectroscopic Surveys*

The Sloan Lens ACS (SLACS) Survey (A. S. Bolton et al. 2004, 2006; T. Treu et al. 2006a; L. V. E. Koopmans et al. 2006; T. Treu et al. 2006b; R. Gavazzi et al. 2007; A. S. Bolton et al. 2008; R. Gavazzi et al. 2008; M. W. Auger et al. 2009, 2010; E. R. Newton et al. 2011; Y. Shu et al. 2015, 2017) selected strong lensing candidates in SDSS spectroscopy via a multifeature background emission-line search. The first SLACS sample (A. S. Bolton et al. 2004, 2006) searched SDSS-I LRGs, targeting [O II] $\lambda\lambda3727$ and at least two of the following: H$\beta$ $\lambda4863$, [O III]$\lambda4960$, and [O III]$\lambda5008$. The confirmed SLACS lens sample[53] includes 93 systems, with an additional five confirmed systems from the SLACS extension (SLACSe; M. W. Auger et al. 2009) and an additional 40 confirmed lenses from the Survey for the Masses (S4TM; Y. Shu et al. 2017). DESI has re-observed eight of these systems as LRGs (six in SLACS, zero in SLACSe, and two in S4TM), two of which (one in SLACS and one in S4TM) are also found among our 4110 candidates. Since SDSS fibers are $3''$ in diameter, twice that of DESI fibers, we do not recover all SLACS lenses.

The BOSS Emission-Line Lens Survey (BELLS; A. S. Bolton et al. 2012; J. R. Brownstein et al. 2012) searched BOSS spectra (observed with the SDSS-III spectroscopic fibers, which have a diameter of $2''$) of all galaxies, regardless of galaxy type, for background emission features from among a list of 11 common strong galaxy lines between 3727 Å and 6733 Å. The confirmed BELLS lens sample[54] includes 36 systems, 20 of which are re-observed as LRGs by DESI. In this work, we identify 11 of these 20 as lens candidates. We recover more candidates that overlap with BELLS than with SLACS because DESI spectroscopic fibers are much closer in size to SDSS-III fibers than to SDSS-I fibers. Later, the BELLS GALaxy-Ly$\alpha$ EmitteR sYstems (BELLS GALLERY; Y. Shu et al. 2016a, 2016b; M. A. Cornachione et al. 2018) searched SDSS-III BOSS spectra for background Ly$\alpha$ $\lambda1216$ emission, which, due to its short emission wavelength, can be observed by SDSS up to much higher background redshifts. None of the 17 confirmed BELLS GALLERY lensing systems[55] were re-observed by DESI as LRGs.

The Spectroscopic Identification of Lensing Objects (SILO; M. S. Talbot et al. 2018, 2021) survey searched spectra observed as part of the Mapping nearby Galaxies at Apache Point Observatory (MaNGA; K. Bundy et al. 2015) survey, part of SDSS-IV, for background [O II] emission lines. SILO implemented similar methods to SLACS and BELLS. Like SDSS-III, the SDSS-IV fibers have a diameter of $2''$. SILO includes 1551 confirmed lenses of different grades,[56] 521 of which DESI re-observed as LRGs. Of these 521,184 overlap with our lens candidates. Because SILO specifically targets [O II], the SDSS-IV fiber size is more similar to DESI's, and the SILO lens sample is much larger than SLACS or BELLS; thus, we find a more significant overlap with our candidates.

[53] Data access: https://simbad.u-strasbg.fr/simbad/sim-ref?querymethod=bib&simbo=on&submit=submit+bibcode&bibcode=2004AJ....127.1860B.

[54] Data access: https://ned.ipac.caltech.edu/uri/NED::InRefcode/2012ApJ...744...41B.

[55] Data access: https://ned.ipac.caltech.edu/uri/NED::InRefcode/2016ApJ...824...86S.

[56] Data access: https://data.sdss.org/sas/dr16/eboss/spectro/lensing/silo/v5_13_0/v5_13_0/1.0. 1/silo_eboss_detections-1.0.1.fits.

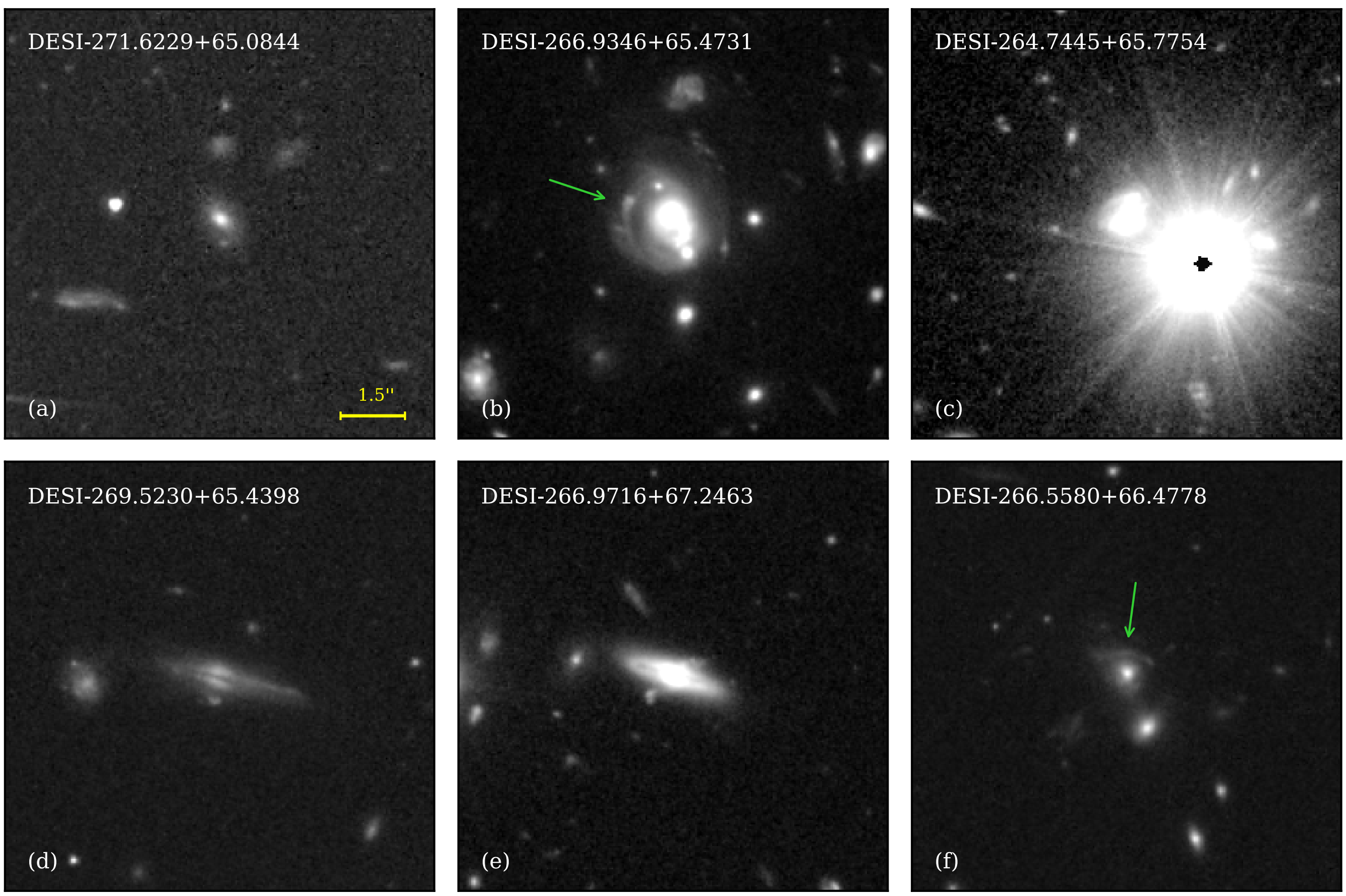

**Figure 12.** Euclid telescope's visible instrument single-broadband filter imaging of six of our 4110 lens candidates, taken from the Euclid Quick Data Release (Q1). Two of these six candidates (panels (b) and (f)) present obvious lensed arcs (that we highlight with green arrows) and are therefore confirmed lenses. Euclid's Strong Lensing Discovery Engine previously discovered the lens in panel (f) (Euclid Collaboration et al. 2025e), but not the lens in panel (b). Imaging of the panel (c) candidate is promising but not obvious enough for confirmation. The signal-to-noise ratio in panel (a) is too low to determine whether the candidate is a lens, and the final two candidates (panels (d) and (e)) are edge-on galaxies and, therefore, require additional signal-to-noise and/or color information for confirmation. The target's name is indicated in the top-left corner, cutouts measure $10'' \times 10''$, and the yellow scale bar indicates the diameter of the DESI fiber ($1\farcs5$).

In total, while DESI has re-observed 534 previously spectroscopically identified lenses as LRGs, only 197 of these 534 objects make it into our catalog. This is likely because SDSS, BOSS, and MaNGA fibers ($3''$, $2''$, and $2''$ diameters, respectively) are larger than DESI fibers ($1\farcs5$ diameter), and can therefore capture light from larger Einstein radius systems. Additionally, differences in precise detection criteria, depth of observations, instrument spectral resolutions, and other effects contribute. A detailed comparison is outside the scope of this work.

In Figure 13, we compare our lens candidates' distribution in $z_{\rm BG}$ versus $z_{\rm FG}$ redshift–redshift space to every previous spectroscopically selected lens sample. The SLACS, SLACS extension, and S4TM lenses are concentrated at both lower foreground and lower background redshifts than our lens candidates. Given the larger spectroscopic fiber size for SDSS-I ($3''$), these samples would select for lower-redshift objects that are more likely to have larger Einstein radii. BELLS and SILO confirmed lenses have background redshifts as high as our sample's, but their foreground redshifts are concentrated around $0.3 \leqslant z_{\rm FG} \leqslant 0.7$, while ours extend up to $z_{\rm FG} = 1.4$. BOSS did not target LRGs past $z = 0.6$. SILO's background redshift distribution goes slightly higher than our sample's due to the fact that both of our searches target the [O II] $\lambda\lambda3727$ doublet, but SDSS-IV's spectrograph observes up to $\lambda = 10,400$ Å while DESI's observing capabilities end at 9800 Å. BELLS GALLERY targets Ly$\alpha$ rather than [O II] and other optical emission lines, resulting in the discovery of much higher-redshift background sources.

In addition to the single-fiber spectroscopic selection method, Y.-M. Hsu et al. (2025)[57] pioneered a new spectroscopic lens search method. If two fibers are placed on targets with sufficient spatial proximity and observe galaxies with a great enough redshift separation, it is likely that the foreground galaxy is acting as a lens to the background galaxy. Following visual inspection, this pair-wise search of DESI spectra yielded 2164 new strong lens candidates in DR1, with 1906 of these being conventional galaxy–galaxy lenses. The other 318 are "dimple lenses" (E. Silver et al. 2025; see their Figure 15), systems with small Einstein radii ($\theta_{\rm E} \leqslant 0\farcs1$) where the foreground galaxy is lower mass than in conventional lensing systems and fainter than the background source. The distortions to the background source in these images are therefore softer and more dimple-like. Note that 28 of our 4110 lens candidates overlap with the pair-wise lens candidates; only one of these is a dimple lens.

### 7.3. *iPTF16geu Re-discovery*

Among our 4110 lens candidates, we recover the host system of the multiply imaged gravitationally lensed type Ia supernova iPTF16geu (A. Goobar et al. 2017). Most lensed transients have been discovered in cluster-scale lenses, making iPTF16geu's discovery in a galaxy-scale lens unique. The Keck Observatory's Near-Infrared Camera 2 with Laser Guide Star Adaptive Optics resolved the four images of the lensed supernova at a distance of $0\farcs3$ from the center of the foreground galaxy. This Einstein radius is much smaller than the $\theta_E$ of most previously discovered lenses, but, as shown in Figure 8, it is in line with the Einstein radii of many of our lens candidates. Our sample should therefore be monitored for future lensed transients. These will enable studies of dark matter substructure in the foreground galaxy via flux ratios between the multiple images of the background source, as well as Hubble constant measurements with time-delay cosmography.

[57] Data access: https://sites.google.com/usfca.edu/neuralens/publications/lens-candidates-hsu-et-al-2025?authuser=0.

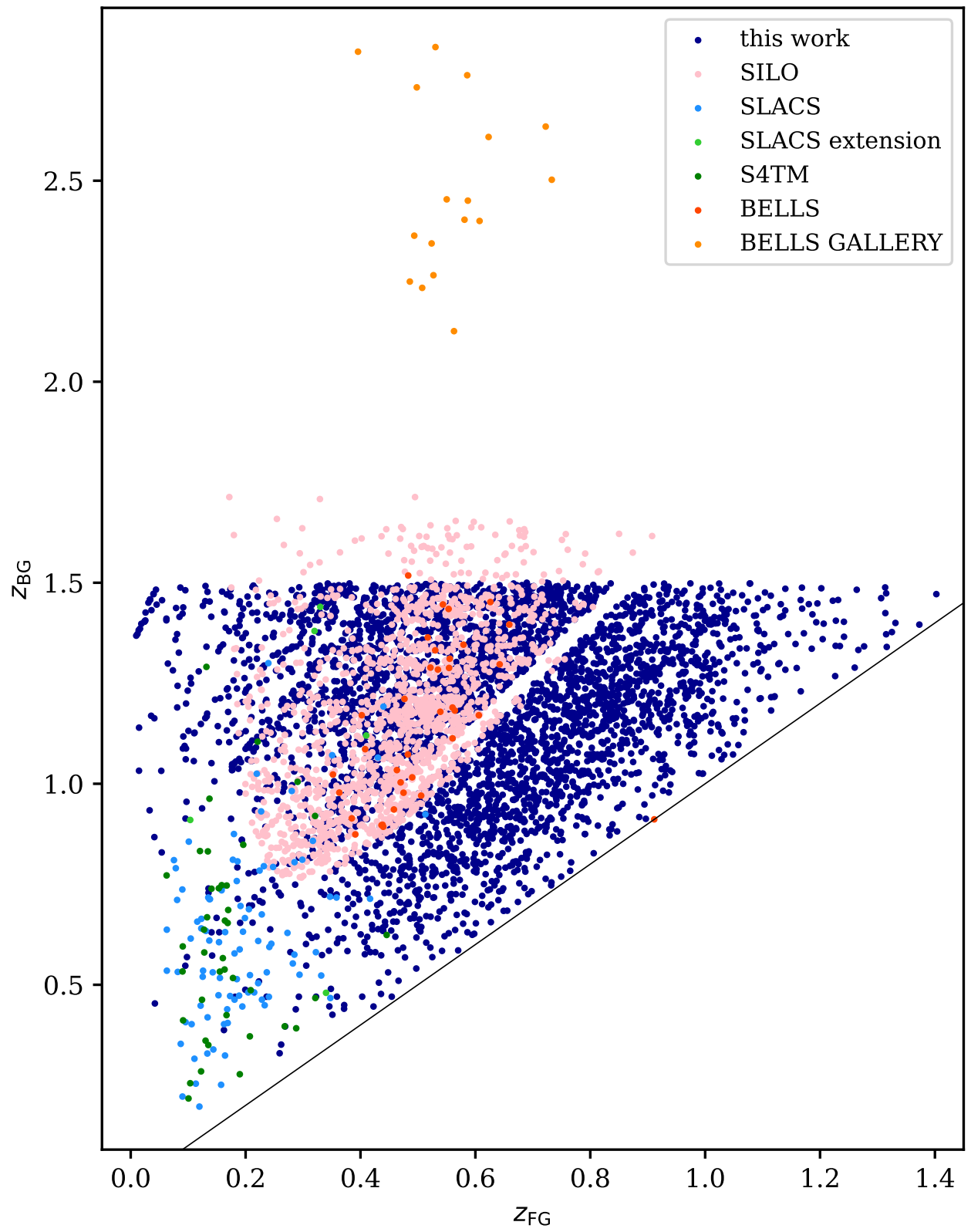

**Figure 13.** Background source redshift ($z_{BG}$) vs. foreground lens redshift ($z_{FG}$) distribution of our spectroscopically identified lens candidates (dark blue) compared to previous samples: SILO (pink), SLACS (blue), SLACSe (light green), S4TM (dark green), BELLS (red), and BELLS GALLERY (orange). The solid black line indicates $z_{FG} = z_{BG}$. Our catalog extends to higher foreground redshifts than all other samples and higher background redshift than SLACS and S4TM. BELLS GALLERY finds lenses with significantly higher $z_{BG}$ because they search for background Ly$\alpha$ $\lambda$1216 emission rather than [O II] $\lambda\lambda$3727.

### 7.4. Double Lens Search

Of particular interest for cosmological studies are double source plane lenses (DSPLs), in which two background sources are being lensed by the same foreground galaxy. Examples include the Jackpot Lens SDSS J0946+1006 (R. Gavazzi et al. 2008), the Eye of Horus HSC J142449-005322 (M. Tanaka et al. 2016), and AGEL150745+052256 (T. M. Barone et al. 2026). The ratio of angular diameter distances between the two lensed sources can be used to constrain the matter density $\Omega_m$ and the dark energy equation of state parameter $w$, independent of $H_0$ (T. E. Collett et al. 2012). To date, only $\mathcal{O}(10)$ DSPLs have been discovered, and only two have been used for such cosmological studies (T. E. Collett & M. W. Auger 2014; N. Sahu et al. 2025). With more DSPL discoveries, population-level studies will significantly improve these constraints.

To search for DSPLs, we reapply the analysis pipeline to the 4110 lens candidates while masking the spectrum within $\pm 200$ km s$^{-1}$ of the previously identified background [O II] $\lambda\lambda$3727 doublet peak wavelength. We do not recover any DSPLs in this manner. There are likely two reasons for this. One, double lens systems are rare. Two, the $\Delta\chi_r^2 > 100$ cut in our selection pipeline ensures that all candidates present an [O II] doublet that is $\sim$10 times better fit by a double Gaussian than any other portion of the residual spectrum. If an LRG was a DSPL whose spectrum presented two [O II] doublets from two distinct background sources, this cut would likely exclude the system from our candidate list entirely.

## 8. Conclusions

We have presented a catalog of 4110 strong galaxy–galaxy gravitational lens candidates. These candidates have been selected from a sample of 5,837,154 LRGs observed by DESI and released as part of the internal Year 3 Data Release, also known as *Loa*, forthcoming as the public DESI DR2. LRGs are massive, intermediate-redshift galaxies that are mostly quenched and have well-defined spectra. We identify lens candidates via the presence of spectral emission features from background sources that may be lensed by the foreground LRG. In particular, we target the singly ionized oxygen doublet [O II] $\lambda\lambda$3727. [O II] is a strong indicator of star formation, and we expect background lensed galaxies at the background redshifts we search to be relatively star-forming. Our strong lens candidates span a wide range of foreground redshifts ($0.01 < z_{FG} < 1.40$) and background redshifts ($0.33 < z_{FG} < 1.50$). Following lensing probability calculations, we expect 2165 of these systems (53%) to be lenses as opposed to chance alignments of non-lensing galaxies.

Because we identify lens candidates spectroscopically, we have precise redshift measurements for both the foreground lens and background source, which is not possible with imaging-based lens discovery. In order to confirm our lens candidates, we require measurements of the spatial alignment of the foreground LRG and background source. We anticipate obtaining such measurements through a combination of data from current and upcoming wide-field imaging surveys (e.g., the Vera Rubin Observatory's Legacy Survey of Space and Time, the Euclid space telescope's Euclid Wide Survey, and the Nancy Grace Roman Space Telescope's planned survey) and targeted follow-up of individual lens candidates that lie outside these surveys' footprints with integral field spectroscopy (e.g., with the Gemini Observatory's Gemini Multi-Object Spectrographs), narrowband imaging, and broadband imaging, which includes the wavelength at which the background [O II] emission feature is identified (e.g., with HST's Advanced Camera for Surveys).

Given that there are only $\mathcal{O}(10^4)$ previously known strong lens candidates, our catalog of 4110 candidates represents a $\sim$50% increase on the literature. Any lenses confirmed from our sample will be valuable for studies of galaxy-scale dark matter distributions at cosmological distances. Any lensed transients observed in our sample can be used to measure dark substructure via flux ratios, as well as to make novel independent measurements of the Hubble constant via time-delay cosmography.

In this study, we have conducted our search only on LRGs. While LRGs are an ideal foreground sample to search for background lensed galaxies due to their large masses and intermediate redshifts, they are not the only type of galaxy that can act as a lens. The *Loa* data release includes a total of 53,937,655 spectra, some of which are stars and quasars, but many of which are galaxies. A simple next step to increase the number of lens candidates we find would be to extend our galaxy–galaxy [O II] emission-line lens search to the DESI emission-line galaxy (ELG) and Bright Galaxy Survey (BGS) samples. BGS is a magnitude-limited survey of ∼10 million galaxies out to $z \simeq 0.2$, and DESI targets ELGs up to $z = 1.6$. Thus, extending our search to these samples will expand the foreground redshift range over which we are able to identify strong lensing candidates. E. McArthur et al. (2025) presented lens candidates from a search of lower-redshift quasars.

Another method by which we can increase our sample of lens candidates is to search for different background emission lines in the spectra of LRGs, ELGs, and BGS galaxies. Single-line searches are not usually very productive, as it is difficult to determine redshift from a single emission line, but the asymmetrical shape of the Ly$\alpha$ 1216 Å line is unique. As demonstrated by the BELLS GALLERY sample, Ly$\alpha$ searches can be quite successful and yield lensed sources with much higher background redshifts than searches targeting [O II] can. Due to the short rest-frame wavelength of the Ly$\alpha$ $\lambda$1216 line, DESI spectrographs can observe Ly$\alpha$ in the redshift range $2 < z < 7$. Other multiline searches can also yield complementary lens candidate catalogs.

## Acknowledgments

This material is based upon work supported by the U.S. Department of Energy (DOE), Office of Science, Office of High-Energy Physics, under contract No. DE-AC02-05CH11231, and by the National Energy Research Scientific Computing Center, a DOE Office of Science User Facility under the same contract. J.K. acknowledges support under the same contract from the U.S. Department of Energy, Office of Science, Office of Workforce Development for Teachers and Scientists (WDTS) under the Science Undergraduate Laboratory Internships (SULI) program. J.K. and N.P. are supported in part by DOE contract No. DE-SC0017660. This material is based upon work supported by the National Science Foundation Graduate Research Fellowship Program under grant No. DGE-2140004.

Additional support for DESI was provided by the U.S. National Science Foundation (NSF), Division of Astronomical Sciences under contract No. AST-0950945 to the NSF's National Optical-Infrared Astronomy Research Laboratory; the Science and Technology Facilities Council of the United Kingdom; the Gordon and Betty Moore Foundation; the Heising-Simons Foundation; the French Alternative Energies and Atomic Energy Commission (CEA); the National Council of Humanities, Science and Technology of Mexico (CONAHCYT); the Ministry of Science, Innovation and Universities of Spain (MICIU/AEI/10.13039/501100011033); and by the DESI Member Institutions: https://www.desi.lbl.gov/collaborating-institutions.

The DESI Legacy Imaging Surveys consist of three individual and complementary projects: the Dark Energy Camera Legacy Survey (DECaLS), the Beijing-Arizona Sky Survey (BASS), and the Mayall z-band Legacy Survey (MzLS). DECaLS, BASS, and MzLS together include data obtained, respectively, at the Blanco telescope, Cerro Tololo Inter-American Observatory, NSF's NOIRLab; the Bok telescope, Steward Observatory, University of Arizona; and the Mayall telescope, Kitt Peak National Observatory, NOIRLab. NOIRLab is operated by the Association of Universities for Research in Astronomy (AURA) under a cooperative agreement with the National Science Foundation. Pipeline processing and analyses of the data were supported by NOIRLab and the Lawrence Berkeley National Laboratory (LBNL). Legacy Surveys also use data products from the Near-Earth Object Wide-field Infrared Survey Explorer (NEOWISE), a project of the Jet Propulsion Laboratory/California Institute of Technology, funded by the National Aeronautics and Space Administration. Legacy Surveys was supported by: the Director, Office of Science, Office of High Energy Physics of the U.S. Department of Energy; the National Energy Research Scientific Computing Center, a DOE Office of Science User Facility; the U.S. National Science Foundation, Division of Astronomical Sciences; the National Astronomical Observatories of China, the Chinese Academy of Sciences and the Chinese National Natural Science Foundation. LBNL is managed by the Regents of the University of California under contract to the U.S. Department of Energy. The complete acknowledgments can be found at https://www.legacysurvey.org/.

Any opinions, findings, and conclusions or recommendations expressed in this material are those of the author(s) and do not necessarily reflect the views of the U.S. National Science Foundation, the U.S. Department of Energy, or any of the listed funding agencies.

The authors are honored to be permitted to conduct scientific research on I'oligam Du'ag (Kitt Peak), a mountain with particular significance to the Tohono O'odham Nation.

## Data Availability

The value-added catalog of lens candidates is publicly available at https://data.desi.lbl.gov/public/papers/gqp/single-fiber-lens/. The same data is also available in Zenodo at DOI: 10.5281/zenodo.17795798.

## ORCID iDs

Juliana S. M. Karp https://orcid.org/0000-0002-1728-8042
David J. Schlegel https://orcid.org/0000-0002-5042-5088
Xiaosheng Huang https://orcid.org/0000-0001-8156-0330
Nikhil Padmanabhan https://orcid.org/0000-0002-2885-8602
Adam S. Bolton https://orcid.org/0000-0002-9836-603X
Christopher J. Storfer https://orcid.org/0000-0002-0385-0014
J. Aguilar https://orcid.org/0000-0003-0822-452X
S. Ahlen https://orcid.org/0000-0001-6098-7247
S. Bailey https://orcid.org/0000-0003-4162-6619
D. Bianchi https://orcid.org/0000-0001-9712-0006
D. Brooks https://orcid.org/0000-0002-8458-5047
F. J. Castander https://orcid.org/0000-0001-7316-4573
A. Cuceu https://orcid.org/0000-0002-2169-0595
A. de la Macorra https://orcid.org/0000-0002-1769-1640
J. Della Costa https://orcid.org/0000-0003-0928-2000
P. Doel https://orcid.org/0000-0002-6397-4457
A. Font-Ribera https://orcid.org/0000-0002-3033-7312
J. E. Forero-Romero https://orcid.org/0000-0002-2890-3725

E. Gaztañaga https://orcid.org/0000-0001-9632-0815
S. Gontcho A Gontcho https://orcid.org/0000-0003-3142-233X
G. Gutierrez https://orcid.org/0000-0003-0825-0517
K. Honscheid https://orcid.org/0000-0002-6550-2023
M. Ishak https://orcid.org/0000-0002-6024-466X
J. Jimenez https://orcid.org/0000-0001-8528-3473
R. Joyce https://orcid.org/0000-0003-0201-5241
S. Juneau https://orcid.org/0000-0002-0000-2394
D. Kirkby https://orcid.org/0000-0002-8828-5463
A. Kremin https://orcid.org/0000-0001-6356-7424
C. Lamman https://orcid.org/0000-0002-6731-9329
M. Landriau https://orcid.org/0000-0003-1838-8528
L. Le Guillou https://orcid.org/0000-0001-7178-8868
M. Manera https://orcid.org/0000-0003-4962-8934
P. Martini https://orcid.org/0000-0002-4279-4182
A. Meisner https://orcid.org/0000-0002-1125-7384
R. Miquel https://orcid.org/0000-0002-6610-4836
J. Moustakas https://orcid.org/0000-0002-2733-4559
S. Nadathur https://orcid.org/0000-0001-9070-3102
W. J. Percival https://orcid.org/0000-0002-0644-5727
C. Poppett https://orcid.org/0000-0003-0512-5489
F. Prada https://orcid.org/0000-0001-7145-8674
I. Pérez-Ràfols https://orcid.org/0000-0001-6979-0125
E. Sanchez https://orcid.org/0000-0002-9646-8198
D. Sprayberry https://orcid.org/0000-0001-7583-6441
G. Tarlé https://orcid.org/0000-0003-1704-0781
R. Zhou https://orcid.org/0000-0001-5381-4372